\documentclass[]{aastex631}
\usepackage{CJK}
\usepackage{graphicx}

\shorttitle{Systematic search for water fountains}
\shortauthors{Fan et al.}
\graphicspath{{./}{figures/}}

\newcommand{\sss}[0]{\scriptscriptstyle}  

\begin{document}
\begin{CJK*}{UTF8}{gkai}

\title{Systematic Search for Water Fountain Candidates using \\the Databases of Circumstellar Maser Sources}

\correspondingauthor{Jun-ichi Nakashima}
\email{junichin@mail.sysu.edu.cn, nakashima.junichi@gmail.com}

\author[0000-0003-3481-5542]{Haichen Fan(范海辰)}
\affiliation{School of Physics and Astronomy, Sun Yat-sen University, Tang Jia Wan, Zhuhai, 519082, P. R. China}

\author[0000-0003-3324-9462]{Jun-ichi Nakashima(中岛淳一)}
\affiliation{School of Physics and Astronomy, Sun Yat-sen University, Tang Jia Wan, Zhuhai, 519082, P. R. China}
\affiliation{CSST Science Center for the Guangdong-Hong Kong-Macau Greater Bay Area, \\Sun Yat-Sen University, 2 Duxue Road, Zhuhai 519082, Guangdong Province, P. R. China}

\author[0000-0001-7102-6660]{D. Engels}
\affiliation{Hamburger Sternwarte, Universit\"{a}t Hamburg, Gojenbergsweg 112, D-21029 Hamburg, Germany}

\author[0000-0002-1086-7922]{Yong Zhang (张泳)}
\affiliation{School of Physics and Astronomy, Sun Yat-sen University, Tang Jia Wan, Zhuhai, 519082, P. R. China}
\affiliation{CSST Science Center for the Guangdong-Hong Kong-Macau Greater Bay Area, \\Sun Yat-Sen University, 2 Duxue Road, Zhuhai 519082, Guangdong Province, P. R. China}

\author[0000-0002-9829-8655]{Jian-Jie Qiu (邱建杰)}
\affiliation{School of Physics and Astronomy, Sun Yat-sen University, Tang Jia Wan, Zhuhai, 519082, P. R. China}
\affiliation{Key Laboratory of Modern Astronomy and Astrophysics (Nanjing University), Ministry of Education, \\Nanjing 210093, Jiangsu Province, P. R. China}
\affiliation{CSST Science Center for the Guangdong-Hong Kong-Macau Greater Bay Area, \\Sun Yat-Sen University, 2 Duxue Road, Zhuhai 519082, Guangdong Province, P. R. China}

\author[0000-0001-7462-5968]{Huan-Xue Feng (冯焕雪)}
\affiliation{School of Physics and Astronomy, Sun Yat-sen University, Tang Jia Wan, Zhuhai, 519082, P. R. China}

\author[0000-0003-1015-2967]{Jia-Yong Xie (谢嘉泳)}
\affiliation{School of Physics and Astronomy, Sun Yat-sen University, Tang Jia Wan, Zhuhai, 519082, P. R. China}

\author[0000-0002-0880-0091]{Hiroshi Imai (今井裕)}
\affiliation{Center for General Education, Institute for Comprehensive Education, \\
Kagoshima University 1-21-30 Korimoto, Kagoshima 890-0065, Japan}
\affiliation{Amanogawa Galaxy Astronomy Research Center (AGARC), Graduate School of Science and Engineering, \\
Kagoshima University 1-21-30 Korimoto, Kagoshima 890-0065, Japan}

\author[0000-0003-2549-3326]{Chih-Hao Hsia (夏志浩)}
\affiliation{The Laboratory for Space Research, Faculty of Science, \\The University of Hong Kong, Cyberport 4, Hong Kong SAR, China}




\begin{abstract}
Water fountains (WFs) are thought to be objects in the morphological evolution of the circumstellar envelopes of low- and intermediate-mass evolved stars, transitioning from spherically symmetric to asymmetric shapes.  We used databases of circumstellar 1612 MHz OH and 22.235 GHz H$_2$O maser sources to search for new WF candidates using the criterion of a larger velocity range of the H$_2$O maser emission compared to that of the OH maser emission. Thus, it is in principle possible to identify WFs with an H$_2$O velocity ranges smaller than those for the previously known WFs. For the OH maser line, we analyzed database entries of 8,474 observations from 2,195 sources, and 6,085 observations from 3,642 sources for H$_2$O maser line. After a close examination of the velocity ranges and line profiles, we identified 11 sources that meet the criterion mentioned above. We examined the IRAS colors of the selected sources and found that two of them (IRAS 19069+0916 and IRAS 19319+2214) are in the color region for post-AGB stars. We find that the maser velocity criterion can discover other astrophysically interesting objects than just WFs. Such objects may include peculiar planetary nebulae with maser emissions and stellar merger remnants.
\end{abstract}



\section{Introduction} \label{sec:Introduction}

Low- and intermediate-mass stars with an initial mass of up to about 8~$\rm{M}_{\odot}$ undergo active mass loss during the Asymptotic Giant Branch (AGB) phase, in which their circumstellar envelopes (CSE) gradually develop with losing material consisting of cold molecular gas and dust to interstellar space \citep[see, e.g.,][]{1996Ap&SS.245..169O,2018A&ARv..26....1H}. In the post-AGB phase, active mass loss ceases and the inner edge of the CSE is detached from the stellar photosphere. One of the notable phenomena that occurs during AGB/post-AGB phases is the morphological change of the CSE, so as to deviate from spherical symmetry and eventually to evolve into many different shapes observed in planetary nebulae (PNe). The physical mechanism of this morphological change is not yet fully understood \citep[see, e.g.,][]{2002ARA&A..40..439B}. Here it must be emphasized that understanding how the AGB/post-AGB CSE changes their shapes is about more than just knowing the dynamics of the CSE: i.e., the structure and kinematics of the CSE are closely related to the molecular chemical reactions that occurred in the CSE.  In order to accurately model these reactions, we need to have a clear understanding of the distributions of temperature, gas/dust density, and internal/external radiation field within CSEs throughout their evolution.

One of the key questions is when the CSE morphological changes begin (in other words, when the CSEs begin to deviate from spherical symmetry). Many previous studies have mentioned that water fountains (WFs) are the key to the problem. \citep[see, e.g.,][]{2009asrp.proc...62I,2013IAUS..290..227I}. WFs are often thought to be low- and intermediate-mass evolved stars that harbor a small-scale collimated molecular jet at the center of their CSEs. The conditions traditionally required to identify WFs are as follows: (1) the velocity range of the 22.235 GHz H$_2$O maser emission (i.e., twice the expansion velocity of the molecular jet) exceeds 100 km s$^{-1}$, (2) the collimated structure of the molecular jet is confirmed by radio interferometric observations, and (3) the infrared properties are not inconsistent with the status of low- and intermediate-mass evolved stars in terms of color index and morphology. To date, 16 WFs are known (see Appendix~\ref{ap:known WF masers}), including several well-studied objects, such as IRAS 16342$-$3814 \citep{1999ApJ...514L.115S}, W43A \citep{2002Natur.417..829I}, IRAS 19134$+$2131 \citep{2004A&A...420..265I}, IRAS 18286$-$0959 \citep{2011ApJ...741...94Y} and IRAS 18043$-$2116 \citep{2023ApJ...948...17U}. Since the length and speed of the molecular jet can be obtained from very long baseline Interferometry (VLBI) observations, the kinematic timescale of the jet can be calculated. The calculated timescales of WF jets have remarkably small values, ranging from a few decades to hundreds of years \citep[see, for example,][]{2003RMxAC..15...20M, 2004A&A...420..265I, 2014A&A...569A..92V, 2020PASJ...72...58I}. However, because maser emission is used as the probe, the derived parameters are a  subject to uncertainty. However, at least for W43A, of which the CO thermal line (CO $J=2-1$) has been observed using  the Atacama Large Millimeter/submillimeter Array (ALMA), it has been confirmed that the kinematic timescale of the molecular jet is definitely short \citep[$\sim 65$ yr;][]{2020ApJ...890L..14T}. In addition, a previous VLBI observation in the 22~GHz H$_2$O maser line \citep{2014A&A...569A..92V} also show that magnetic fields are involved in jet formation. On the base of the number of known WFs in the Milky Way and the assumed duration of a WF jet, \citet{2022NatAs...6..275K} recently  proposed that WF sources are binary stars which are  just experiencing a common-envelope phase.

An important issue when considering the relationship between the evolution of low- and intermediate-mass evolved stars and the formation of WFs is to accurately determine the current evolutionary state of the known WFs.  Considering the observational properties of the known WFs, the evolution of the CSE to deviate from spherical symmetry may begin earlier than   the post-AGB phase. For example, SiO masers were detected in some WFs \citep[see, e.g.,][]{2003PASJ...55..229N,2022AJ....163...85A}. Most SiO maser sources are found around AGB stars or red supergiants, with several known exceptions for young stellar objects  \citep[YSOs; see, e.g.,][]{2016ApJ...826..157C}. Even if there exists a period of emitting the SiO maser after the transition to the post-AGB phase, it is likely to be short, and very close to the AGB phase.  Conversely, some WFs have been proposed as more evolved objects, such as PNe \citep{2015ApJ...799..186G}. If the number of known WF samples is large, it would be possible to discuss the evolutionary status of WFs from a statistical analysis, in terms of color index, etc., for instance, using public infrared photometric data. 

The WFs known so far were mostly found by their unusually wide velocity range ($>$ 100 km s$^{-1}$) of H$_2$O maser emission. However, the apparent velocity of the WF jet is expected to vary with the inclination angle and the degree of jet development, i.e. how much time has elapsed since the acceleration began \citep{2002A&A...388..252E, 2013ApJ...769...20Y}. Therefore, the sample bias caused by selecting objects by a physically unjustified jet velocity criterion (i.e., $>$ 100 km s$^{-1}$) would be another serious scientific problem.

In view of these two problems, our team used for the first time the comparison of the H$_2$O and OH maser velocity range systematically to find new WF candidates \citep{2013ApJ...769...20Y}. In the present study, with the aim of searching for new WFs, we cross-checked H$_2$O (22.235 GHz) and OH (1612 MHz) maser sources with the databases of circumstellar maser sources and also performed a large-scale systematic comparison of the velocity ranges of H$_2$O and OH maser lines. We then investigated the basic properties of the sources satisfying the velocity conditions proposed in \citet{2013ApJ...769...20Y} using maser spectra from the literature and infrared archival data. Finally, we discuss how effective such a comparison of maser line velocity ranges is as a method of selecting WF candidates, and what other types of objects might be selected under the same selection conditions. Because a detailed explanation of the methodology is not provided in \citet{2013ApJ...769...20Y}, it is given in Section~\ref{sec:Methodology} in this paper.

\section{Methodology and Data Processing} \label{sec:Methodology}

\subsection{Basic ideas for finding WF candidates using circumstellar maser source databases} \label{subsec:basic}

Firstly we briefly summarize the basic ideas behind our WF candidate search, and in the following subsections we describe the more specific search process in detail. The main purpose of the present work is to select WF ``candidates'', and therefore confirmatory interferometric observations are needed to conclude that each of them is indeed a WF. The limitations of our search and the properties of the selected objects are discussed in Section~\ref{sec:Discussions}.

\begin{figure}[h]
    \centering
    \includegraphics[width = 0.7\textwidth]{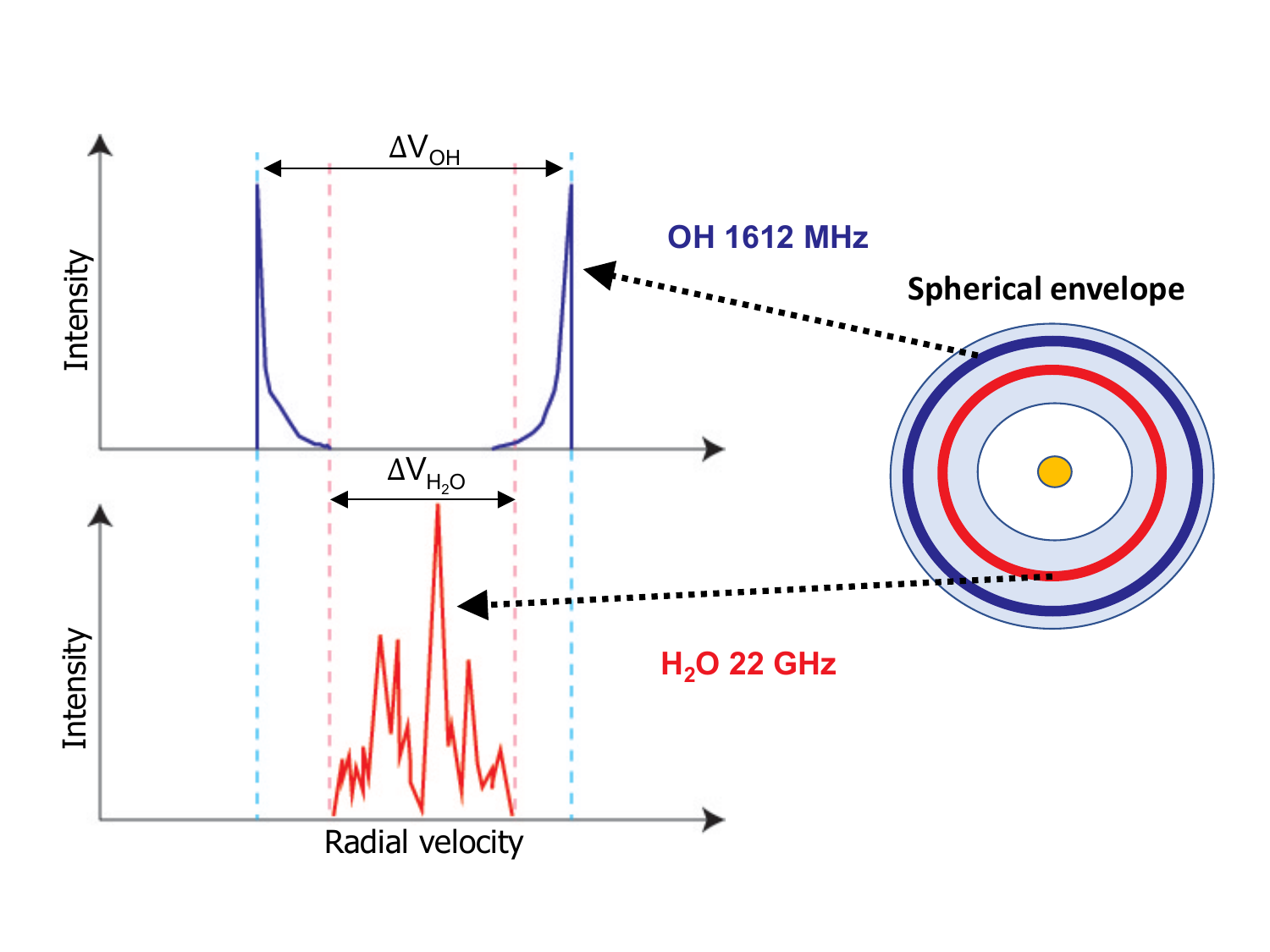}
    \caption{Schematic view of the spherically expanding CSE emitting the OH 1612~MHz and H$_2$O 22.235~GHz maser lines. The right panel shows the structure of the CSE, and the left panel shows the conceptual line profiles of the OH and H$_2$O maser lines. The emission regions of the OH and H$_2$O masers within the CSE are shown in the same color as the spectra (i.e. dark blue for OH and red for H$_2$O masers). $\Delta$V$_{\rm OH}$ represents the velocity difference between the red-shifted and blue-shifted peaks of the OH maser line. $\Delta$V$_{\rm H_2O}$ is the velocity range of the H$_2$O maser emission.}
    \label{fig:idea-spherical}
\end{figure}

The working hypothesis here is that a CSE is fully spherical before a WF jet is launched. This situation is illustrated in Figure~\ref{fig:idea-spherical}. A spherically expanding CSE of an evolved star with oxygen-rich chemistry often emits H$_2$O maser and OH maser lines (note:``oxygen-rich''  means that the number of oxygen atoms exceeds that of carbon atoms). Here we suppose a CSE in which both maser lines are detected. Due to the different physical conditions required to excite the maser lines, the location of the H$_2$O maser emission region is closer to the central star than the OH maser emission region \citep[generally, the H$_2$O maser line is emitted near the central star within a few tens of stellar radii, and the OH  maser line is emitted from the region within a few hundred stellar radii; see e.g, ][]{2011A&A...525A..56R,2017AJ....153..119O}. In addition, the formation of OH from H$_2$O molecules by photodissociation due to interstellar UV radiation also affect the spatial distribution of H$_2$O and OH. In the H$_2$O maser region, the molecular gas is accelerated outwards by the stellar radiation pressure on the dust component \citep[stellar pulsation could also be involved in the acceleration mechanism; see, e.g.,][]{2018A&ARv..26....1H}. As a result, there is no fixed pattern to the H$_2$O maser line profile, and it can take on many different types of profiles. On the other hand, in the OH maser region, the molecular gas has already reached the terminal velocity. Thus, the line profiles of the OH maser emission of the spherically expanding envelope are known to be in the form of double peaks corresponding to the receding and approaching sides of the CSE \citep[see, e.g.,][]{1996A&ARv...7...97H}.  Importantly, in the spherically expanding envelope, the radial velocity of the H$_2$O maser emission never exceeds the velocity range of the OH maser emission.  $\Delta V_{\rm OH} \geq \Delta V_{\rm H_{2}O}$ always holds for a spherically expanding envelope.

\begin{figure}[h]
    \centering
    \includegraphics[width = 0.7\textwidth]{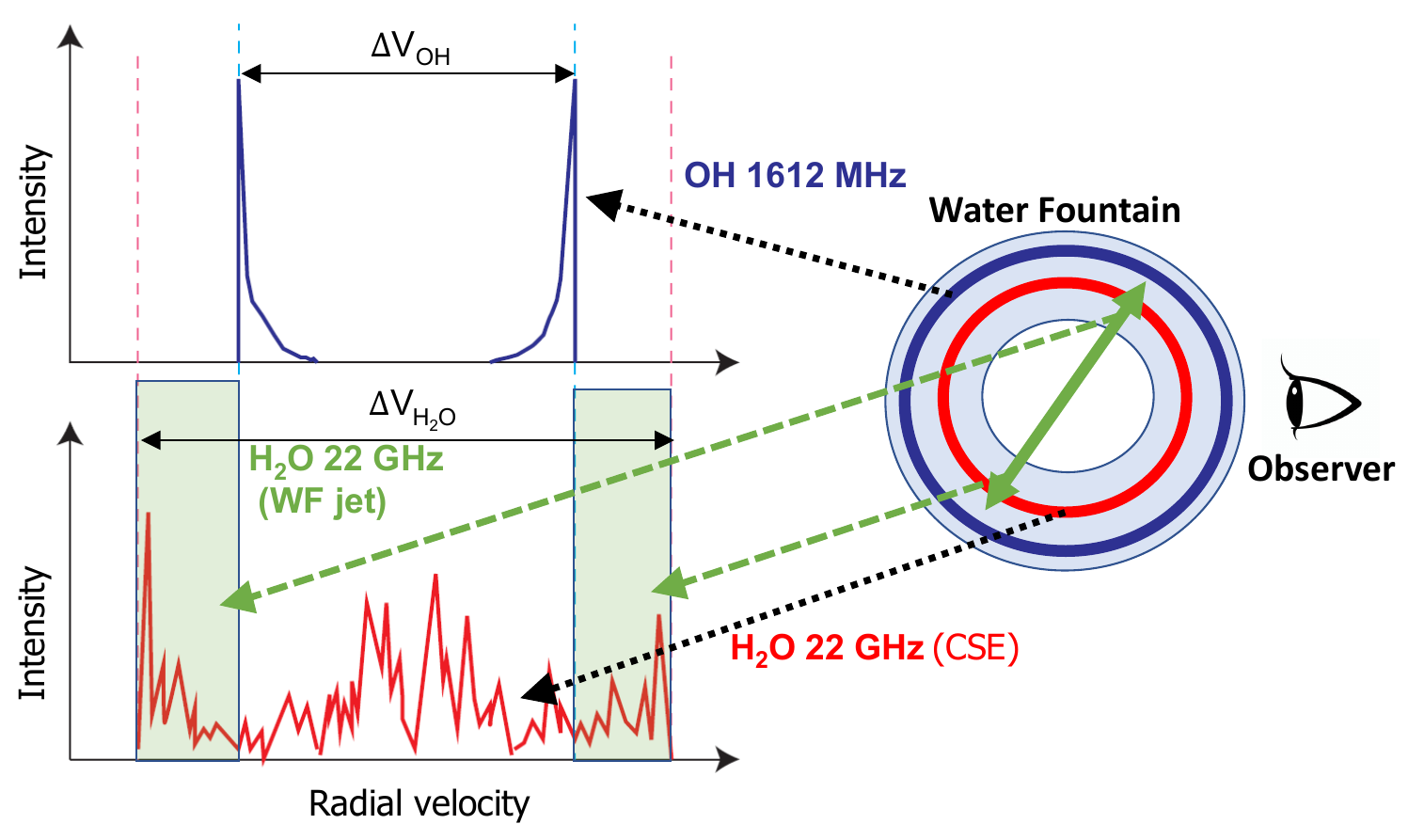}
    \caption{Schematic of the CSE of a water fountain emitting the OH 1612~MHz and H$_2$O 22.235~GHz maser lines. It differs from Figure~\ref{fig:idea-spherical} in the presence of a WF jet at the center of the spherically expanding envelope. The two-sided green arrow denotes  the WF jet that produces double-peaked spectral profile. Note that there is still no conclusive information on the relative size of the WF jet to the spherically expanding envelope, and there are uncertainties in the representation. There also may exist low-velocity components, which may be associated with the WF jet and/or relics of a spherical CSE, within the width represented by $\Delta$V$_{\rm OH}$ in the left panel.  In this figure, it is assumed that the velocity and inclination of the jet with respect to the observer's line of sight combine to produce a H$_2$O velocity range that is larger than that of the OH emission.}
    \label{fig:idea-wf}
\end{figure}

Secondly, we consider the situation after the launch of a WF jet hosting H$_2$O maser emission. A schematic of this situation is illustrated in Figure~\ref{fig:idea-wf}. Here, we assume that the expansion velocity of the WF jet is faster than that of the spherically expanding envelope. Under this assumption, if the jet axis is not nearly perpendicular to the line-of-sight direction, then the high-velocity component of the H$_2$O maser emission emitted by the WF jet is detected outside the OH velocity range (namely, $\Delta V_{\rm OH} \leq \Delta V_{\rm H_{2}O}$). This means that for CSEs in which both the H$_2$O 22.235~GHz and the OH 1612~GHz maser lines are detected, if $\Delta V_{\rm OH} \leq \Delta V_{\rm H_{2}O}$ relation is satisfied, then such a maser source can be considered a candidate for WF.


\subsection{Data used in the analysis} \label{subsec:data}

\begin{figure}[h]
    \centering
    \includegraphics[width = 1.0\textwidth]{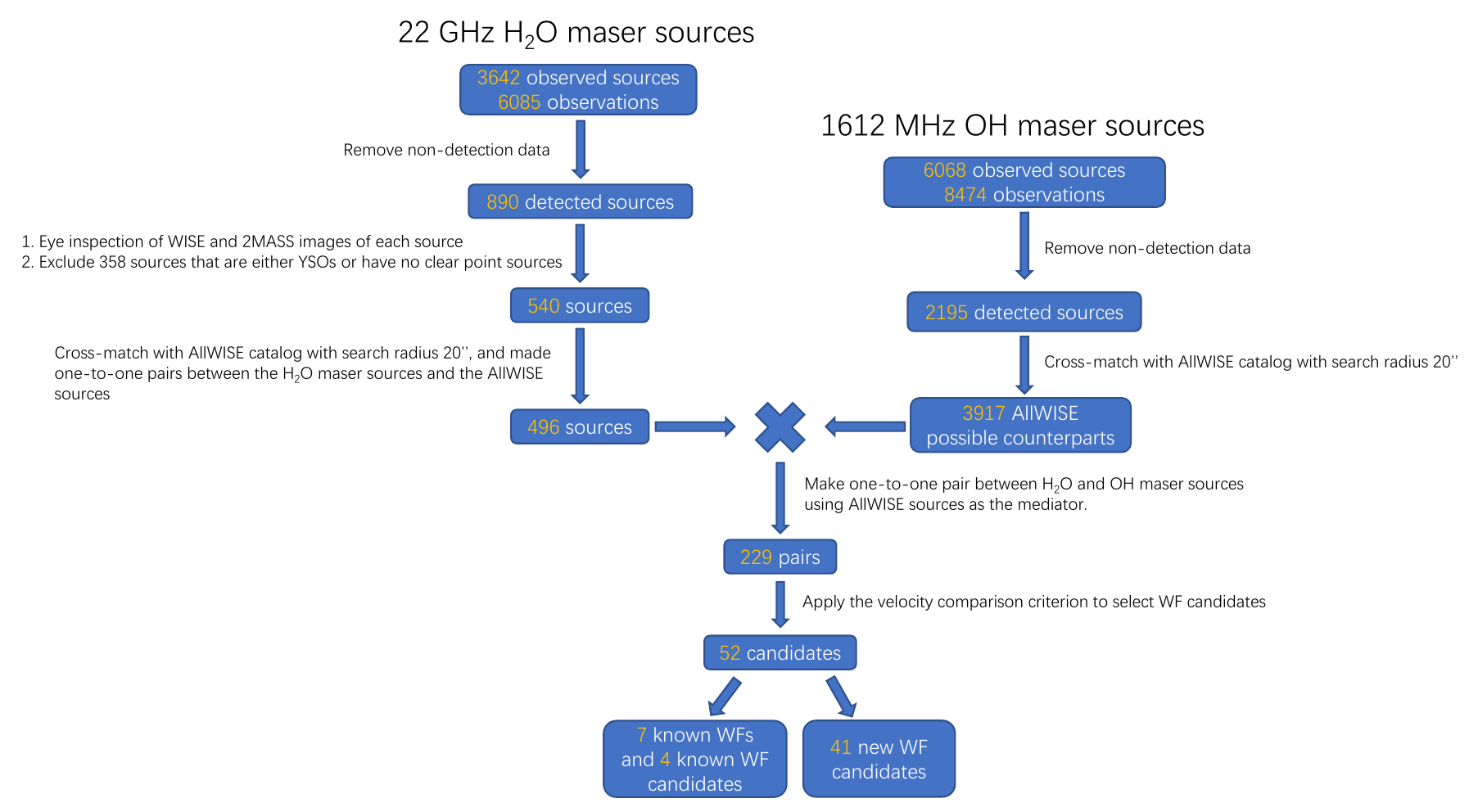}
    \caption{Flowchart of the source selection in this study.}
    \label{fig:flowchart}
\end{figure}

Figure~\ref{fig:flowchart} shows the overall flow of the data analysis performed in the present study (excluding the process of more detailed inspection discussed in Section~\ref{subsec:indiv}). As a source of information on OH maser sources, we used the database of \citet[][hereafter OH database]{2015A&A...582A..68E} accessible from CDS/VizieR\footnote{https://cdsarc.cds.unistra.fr/viz-bin/cat/J/A+A/582/A68}. This database contains the records of 13,655 observations toward 2,341 sources in the OH 1612 MHz, 1665 MHz, and 1667 MHz lines. In the present analysis we used only the 1612 MHz line data, which consist of 8,474 observations toward 6,068 sources (2,195 out of them are detected sources). For the H$_2$O 22.235 GHz maser line, we used an unpublished database (hereafter H$_2$O database) compiled by the same authors. The version 0.1 of the database used for our analysis contains records of 6,085 observations towards 3,642 sources (890 of them are detected sources). A machine-readable table listing the detected H$_2$O maser sources and the reference information of the H$_2$O maser observations is appended to this paper (see, Sections~ \ref{subsec:h2o-catalog} and Appendix~\ref{ap:h2o-catalog}). The H$_2$O maser database contains data from 57 papers. It should be noted that these searches are based on different selection criteria (including blind surveys), have different sensitivity limits, cover different areas of the sky, etc., and therefore the database must be considered heterogeneous in nature. Maps with the distributions of the OH and H$_2$O maser sources in the sky are given in the appendix~\ref{ap:distribution} (see, Figures~\ref{fig:OH-distribution} and \ref{fig:H2O-distribution}).

In addition to the above two databases of circumstellar maser sources, three infrared data archives were used: i.e., the AllWISE Data Release\footnote{https://wise2.ipac.caltech.edu/docs/release/allwise/} of the Wide-field Infrared Survey Explorer\footnote{https://wise2.ipac.caltech.edu/docs/release/allsky/} \citep[WISE,][W1 (3.4~$\mu$m), W2 (4.6~$\mu$m), W3 (12~$\mu$m) and W4 (22~$\mu$m)]{2010AJ....140.1868W, 2011ApJ...731...53M}, the Two Micron All Sky Survey\footnote{https://irsa.ipac.caltech.edu/Missions/2mass.html} \citep[2MASS,][J (1.235~$\mu$m), H (1.662~$\mu$m) and K$_s$ (2.159~$\mu$m) ]{2006AJ....131.1163S},
IRAS catalog of Point Sources, Version 2.0\footnote{https://cdsarc.cds.unistra.fr/viz-bin/cat/II/125} (IRAS PSC, 12, 25, 60 and 100~$\mu$m).

The AllWISE data were used for two purposes: (1) to know the exact location of the maser sources by identifying the WISE counterparts, and (2) to remove contamination, especially YSOs, according to the infrared morphology and color indices. The 2MASS data were originally intended to be used for the same purpose as the AllWISE data. However, since the observing wavelengths of 2MASS are shorter than those of WISE, there exist many cases where counterparts of maser sources are not found, so the 2MASS data were used only as an additional aid to remove contamination by inspecting the near-infrared morphology. The IRAS PSC was primarily used to compare the mid-infrared properties of the selected sources with previous studies.

\subsection{Processing of H$_2$O maser data and cross-checking with infrared sources} \label{subsec:h2o-catalog}

The database of circumstellar H$_2$O masers contains the following information: i.e., name of the source, coordinate values used in the observation, date of the observation, the maximum and minimum velocities at which maser emission was detected, the velocity at the emission peak, velocity integrated intensity of the maser emission, root-mean-square noise level, and bibliographic information of the observation. This database includes non-detection records, and the number of objects for which H$_2$O maser emission has been detected is 890 out of 3,642 objects\footnote{For these 890 sources, a machine-readable table summarizing the coordinate information used in the radio observations, the coordinate information of the WISE counterparts, the results of the infrared image inspection, and bibliographic information is attached to this paper. See Appendix~\ref{ap:h2o-catalog} for details.}. These 890 H$_2$O maser sources include all 16 known WF sources and 4 WF candidates found by \citet{2013ApJ...769...20Y, 2014ApJ...794...81Y}\footnote{\citet{2013ApJ...769...20Y, 2014ApJ...794...81Y} actually found five WF candidates, one of which (IRAS 18455+0448) was confirmed for WF status by \citet{2014A&A...569A..92V}}. 

The role of the H$_2$O maser line in this analysis is to trace the high-velocity component possibly caused by the WF jet, as mentioned in Section~\ref{subsec:basic}. The maximum and minimum velocities for each observation are included in the database. Since the maser line profile varies with time, when multiple observations were available for a single object, the maximum and minimum velocities were determined by comparing all observations. In the following text we will refer to the maximum and minimum velocities of the H$_2$O maser emission as $V_{\rm max}$ and $V_{\rm min}$ respectively.

\begin{figure}[h]
    \centering
    \includegraphics[width = 1.0\textwidth]{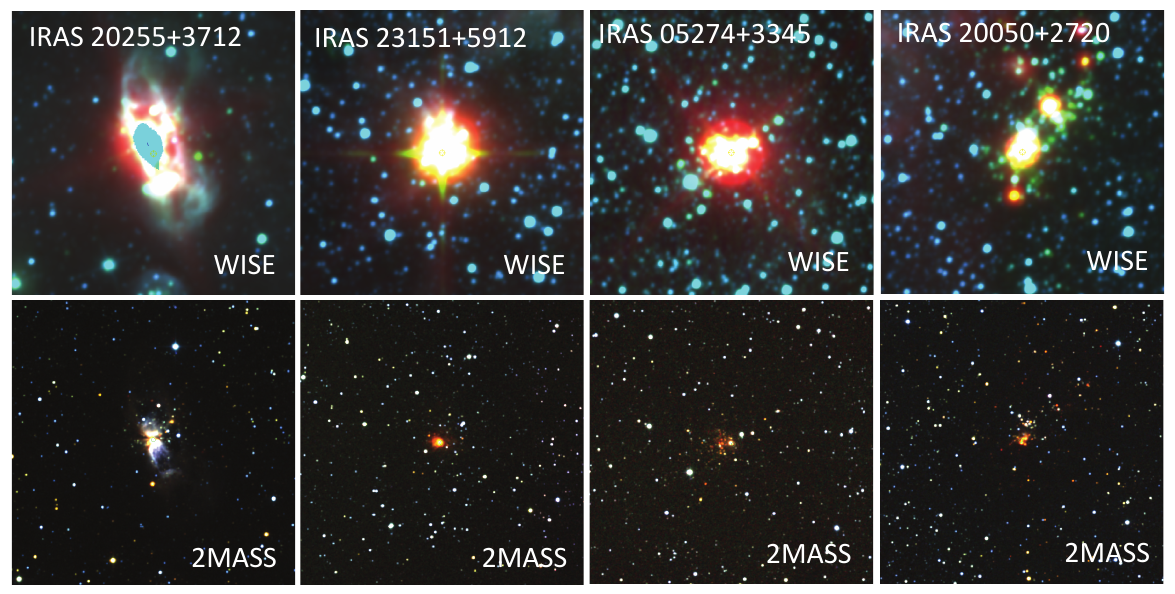}
    \caption{Examples of sources excluded by eye inspection of infrared images. The top row shows composite color images created from WISE W1, W2, and W4 bands. The bottom row shows composite color images created from 2MASS J, H, and K bands. For both WISE and 2MASS images, the wavelength bands are assigned to the blue, green, and red channels in order from shortest to longest wavelength when the composite color image is generated. Both image sizes are 600$''$ square, and the north is up, east is left. The two sources on the left were excluded as obvious YSOs. The two on the right were excluded because they could not be constrained to a single infrared counterpart to the maser source (these sources also appear to be YSOs).}
    \label{fig:contamination}
\end{figure}

Here, it should be noted that the database is an early version that is still under construction, and that the database may contain a significant number of YSOs. Therefore, initially, for a subset of the above 890 objects with H$_2$O maser emission, we briefly examined WISE and 2MASS infrared images to see how serious the YSO contamination is. Specifically, we checked 2MASS and WISE images for the presence of extended nebulosity, and indeed we found a non-negligible number of objects with a strong possibility of being YSOs (see, Figure~\ref{fig:contamination}). Then, based on the result of this initial inspection, we further inspected the WISE and 2MASS images by eye for all 890 objects and identified 128 YSO candidates with extended nebulosity. This eye inspection was done using the Aladin Sky Atlas\footnote{http://aladin.cds.unistra.fr/aladin.gml} provided by CDS on the web. The data of these YSO candidates were excluded from further analysis. The results of the inspection are summarized in the machine readable table (see, Appendix~\ref{ap:h2o-catalog}). It is also true that there are cases where it is difficult to distinguish between YSOs and evolved stars based solely on the eye inspection of IR images \citep{2012A&A...545A..20R}. However, such cases where the distinction is difficult are rare, and in most cases the nebulosity of AGB/post-AGB seen in IR images can basically be distinguished from YSOs by the shape and extent of the nebulosity.

Then additional 230 H$_2$O maser sources were eliminated from the WF candidates similarly by eye inspection. The eliminated sources encounter the following two cases: (1) no clear infrared counterpart is found within about 1$'$ of the H$_2$O maser source, and (2) the location of the maser source is close to the Galactic Center or the Galactic plane that are too crowded with infrared sources for identifying the counterpart. These maser sources were excluded from the analysis for two reasons. First, their property is not consistent with that of the known WF sources that have bright infrared counterparts. Second, if the source is in a crowded region, it is not possible to fully constrain the infrared counterpart due to the lack of angular resolution of radio observations.

We have noticed that among the known WFs there exists a case where the position of the H$_2$O maser source and that of the infrared counterparts given in the infrared source catalog do not match. For example, IRAS 18139$-$1816 is such a case. The position of the WISE and 2MASS infrared point source occasionally differs by about 30$''$ from the coordinates recorded in the database of H$_2$O maser sources. The misalignment is caused by the fact that early H$_2$O maser observations were made using IRAS positions, which can sometimes have an uncertainty larger than 30$''$ (see, Section~\ref{subsec:pairs}). We have excluded such cases to ensure the reliability of the analysis.

\begin{figure}[h]
    \centering
    \includegraphics[width = 0.6\textwidth]{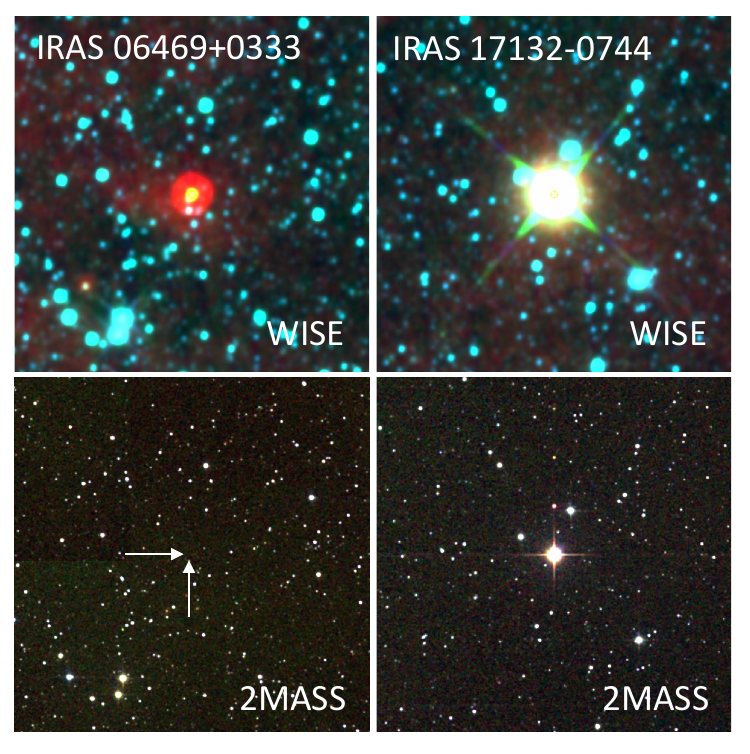}
    \caption{Examples of sources that could be clearly identified as counterparts to the maser source. The details of the infrared images are the same as in Figure~\ref{fig:contamination}. The source on the left (IRAS 06469+0333) is an example of a clear counterpart found only in the WISE images (see text for details). The location of the H$_2$O maser source is around the tip of the white arrows in the lower left panel (2MASS image), but no corresponding red point source is clearly visible in this vicinity.}
    \label{fig:point-source}
\end{figure}

Then we identified  infrared stellar counterparts (i.e., infrared point sources) for the remaining H$_2$O maser sources using the WISE or 2MASS catalogs to obtain accurate coordinate information. We found that all of those H$_2$O maser sources have WISE counterparts in at least one of the 4 WISE bands, but many sources were not identified in the 2MASS J, H or K images (see Figure~\ref{fig:point-source}). Therefore, we cross-checked the WISE point sources with the maser sources.

To perform this cross-matching, we used the X-Match\footnote{http://cdsxmatch.u-strasbg.fr/} service provided by CDS. When the cross-matching radius was set to 20$''$ and the H$_2$O maser sources were matched against the AllWISE catalog, about 10\% of the H$_2$O maser sources had two or more WISE sources matched. In these cases, we checked the WISE infrared colors to determine the correct counterpart. Mass-losing evolved stars are typically surrounded by a cold dust envelope with an effective temperature of about 150--300 K. As a result, their spectral energy distribution (SED) has an excess at mid-infrared wavelengths. Therefore, the AllWISE colors  can be used to exclude sources other than evolved stars. We determined the search radius in X-Match to be 20$''$ after much trial and error. According to our eye inspection there exist 11 sources that have point source counterparts in the WISE images, but for which no counterpart was found in the AllWISE catalog. These objects were excluded from the subsequent analyses for reliability reasons.

We calculated color indices using a combination of the W1 and W4 fluxes and the W2 and W4 fluxes of the WISE source to see if there exist a mid-infrared excess in the possible counterpart. Since the flux values in the AllWISE catalog are given with respect to Vega magnitudes, a color index value of 0 does not correspond to equal flux in the two bands. Therefore, we first converted the WISE magnitudes to fluxes in unit of Jy as described in \citet{2010AJ....140.1868W} and then calculated the color indices as follows: [W1]-[W4]=$2.5\log(F_{\sss\text{W4}}/F_{\sss\text{W1}})$ or [W2]-[W4]=$2.5\log(F_{\sss\text{W4}}/F_{\sss\text{W2}}$). Using these color indices, we eliminated WISE point sources with no infrared excess (i.e., [W1]-[W4]$<0$ and [W2]-[W4]$<0$) from the candidates for counterparts. The WISE photometric values for some sources have a low quality flag due to saturation, but the saturation is not a serious problem here since the WISE photometric values were used to check for excess intensity in the W4 band over the W2 band (in evolved stars, the difference in intensity between W2 and W4 is quite obvious). There exist a very small number of cases where the counterparts could not be constrained to a single object based on the color index alone (i.e., there are two red sources within a beam). In such a case, priority was given to those with high intensity in the W4 band. When there are two or more sources of the same color and intensity, the choice would be difficult, but fortunately such a case did not exist in the present samples. Through the above process, we finally uniquely identified WISE point sources that are considered mass-losing evolved stars for 496 H$_2$O maser sources.

\subsection{Processing of OH maser data and velocity comparison} \label{subsec:oh-catalog}

The 1612 MHz OH maser database of \citet{2015A&A...582A..68E} contains information such as source name, coordinate values, bibliographic information of the observations, velocity information of the line, line intensity, and profile classification for each source. In this analysis we used the velocities of the blue-shifted and red-shifted peaks of the double peak profile of the OH maser line associated with a spherically expanding CSE. Hereafter, these velocities are denoted as $V_b$ and $V_r$. However, at times only a single peak (either on the receding or on the approaching side of the spherically expanding envelope) is detected due to time variability of the maser emission and/or inhomogeneous gas distribution. In such cases the  peak velocity is recorded in the $V_b$ column by default, unless the peak can be assigned to the blue-  or red-shifted part of the OH maser profile, using other independent observations.

Similar to the H$_2$O maser database, there are many cases where multiple observations are registered for a single object. In such cases, to obtain the most accurate expansion velocity of the spherically expanding envelope, we chose the largest value of all previous observations for $V_r$ and the smallest value of all previous observations for $V_b$ as the representative value. Observations for which no values were given for the $V_r$ and $V_b$ columns (i.e. non-detection) were excluded from the analysis.

We then cross-matched the OH maser sources with the AllWISE point sources using the X-Match service, in the same way as for the H$_2$O maser sources. As with the H$_2$O maser data, the search radius was set to 20$''$. Then, if an AllWISE source corresponding to the H$_2$O maser source, which is determined in Section~\ref{subsec:h2o-catalog}, was found within 20$''$ of the OH maser source, it was assigned to be the counterpart as the OH maser source. That is, the AllWISE source acts as an interface to pair the OH maser source with the H$_2$O maser source.

Using the above procedures, we uniquely matched 229 objects between OH maser , H$_2$O maser , and AllWISE sources. For these objects we performed velocity comparisons as described in Section~\ref{subsec:basic}. Specifically, for the case of a double peak profile of the OH maser source, we compared the velocity range of the H$_2$O emission ($V_{\rm min} < V_{\rm lsr} < V_{\rm max}$) with that of the OH maser emission ($V_b < V_{\rm lsr} < V_r$). We have supposed that the maximum and minimum velocities of the H$_2$O maser emission ($V_{\rm max}$ and $V_{\rm min}$) need not both exceed the velocity range of the OH maser emission to allow for the possibility of time variability in the intensity of the H$_2$O maser emission (or to deal with uncertainties caused by inhomogeneity of the material distribution) for identification of a WF candidate.

We also consider the case where the OH maser has a single peak. In this case, the velocity of the single peak is denoted as $V_s$ ($= V_r = V_b$). If either the maximum or minimum velocity of the H$_2$O maser emission (i.e., $V_{\rm max}$ or $V_{\rm min}$) is more than 60~km~s$^{-1}$ away from the peak velocity of the OH maser emission (i.e., $V_s$), the object is also considered as WF candidate. This means that we set an upper limit of 30~km~s$^{-1}$ for the expansion velocity of the spherically expanding envelope formed during the AGB phase. It is generally known that the expansion velocity of the CSE of AGB stars is usually between 10--20~km~s$^{-1}$, but we set 30~km~s$^{-1}$ here to ensure a more reliable selection of WF candidates.

After comparing the velocities, 52 objects met the criteria. Of these 52, seven were already known WFs and four were previously reported WF candidates (see, Section~\ref{subsec:confirmation} for known WFs in the sample). This means that 41 of the 52 objects are ``potentially'' new WF candidates. However, as mentioned in Section~\ref{subsec:basic}, we note again that we cannot immediately say that these 41 are WF candidates as they are. The properties of these 41 objects will be carefully discussed in more detail in following sections.

\section{Results} \label{sec:Results}
\subsection{Overview of the selected sources} \label{subsec:overview}

\begin{figure}[h]
    \centering
    \includegraphics[width = 0.9\textwidth]{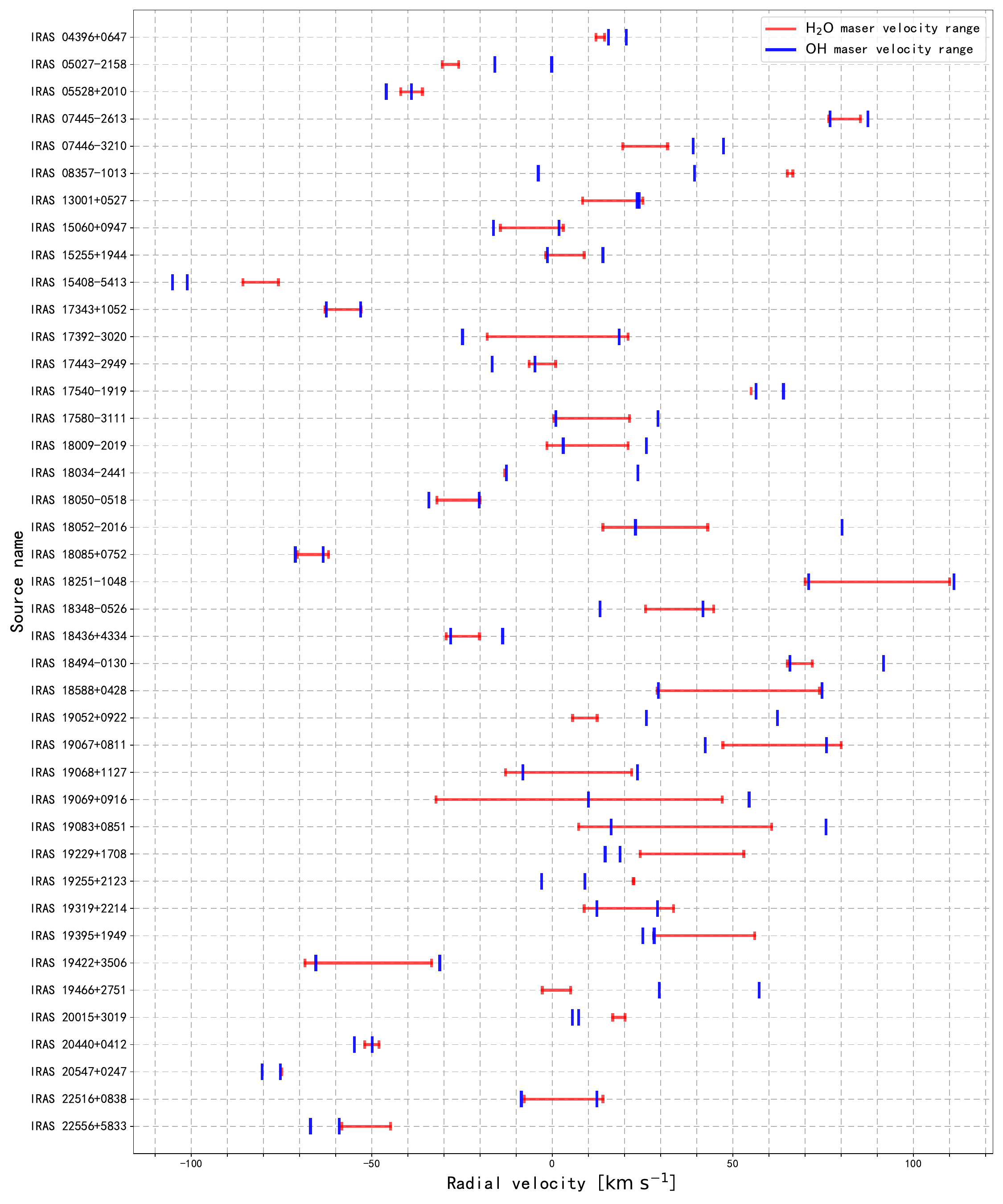}
    \caption{Comparison of the velocity ranges of the H$_2$O 22.235 GHz and OH 1612 MHz maser lines for 41 selected sources. The blue vertical lines represent the velocities of the red-shifted and blue-shifted peaks of the OH maser line. The red line represents the velocity range of the H$_2$O maser emission. The sources are arranged in ascending order with respect to right ascension.}
    \label{fig:vel-comp}
\end{figure}

Finally, we found 41 OH/H$_2$O maser sources that satisfy the velocity conditions described in Section~\ref{subsec:basic}. This means that 17.9\% of the 229 sources meets the velocity condition. The maser velocities, coordinate information (WISE position), and infrared (IRAS) fluxes for the infrared counterparts of the 41 selected sources are summarized in Tables~\ref{tab:velocity} and \ref{tab:infrared}. The results of the velocity comparison between the OH and H$_2$O maser lines are shown in Figure~\ref{fig:vel-comp}. In the past, H$_2$O maser sources with velocities greater than 100~km~s$^{-1}$ were identified as WFs. However, this search did not find any such ``classical" WFs that were slipped from previous research. (Some of the sources classified as YSO candidates have H$_2$O maser velocity ranges exceeding 100~km~s$^{-1}$. These sources are summarized in Table~\ref{tab:YSOs} in Appendix~\ref{ap:YSO} .) Half the velocity between the two peaks of the OH maser emitted by the spherically expanding envelope corresponds to the expansion velocity of the envelope. This expansion velocity is typically about 10--20~km~s$^{-1}$, which corresponds to 20--40~km~s$^{-1}$ in the velocity range shown in Figure~\ref{fig:vel-comp}. For some sources the velocity range of the OH maser emission is clearly less than 10~km~s$^{-1}$. This can be explained by two possible scenarios: (1) the expansion velocity of the spherically expanding envelope is slow, or (2) the expansion velocity of the spherically expanding envelope is within the normal range, but only one peak is detected, either on the receding or on the approaching side of a spherically expanding envelope. By examining the OH maser line profiles of individual sources, it is possible to make some speculations as to which of the two situations applies. A detailed discussion of maser profiles for individual sources is given in Section~\ref{subsec:indiv} and Appendix~\ref{ap:indiv} and \ref{ap:other}.

\begin{figure}
    \centering
    \includegraphics[width = 0.7\textwidth]{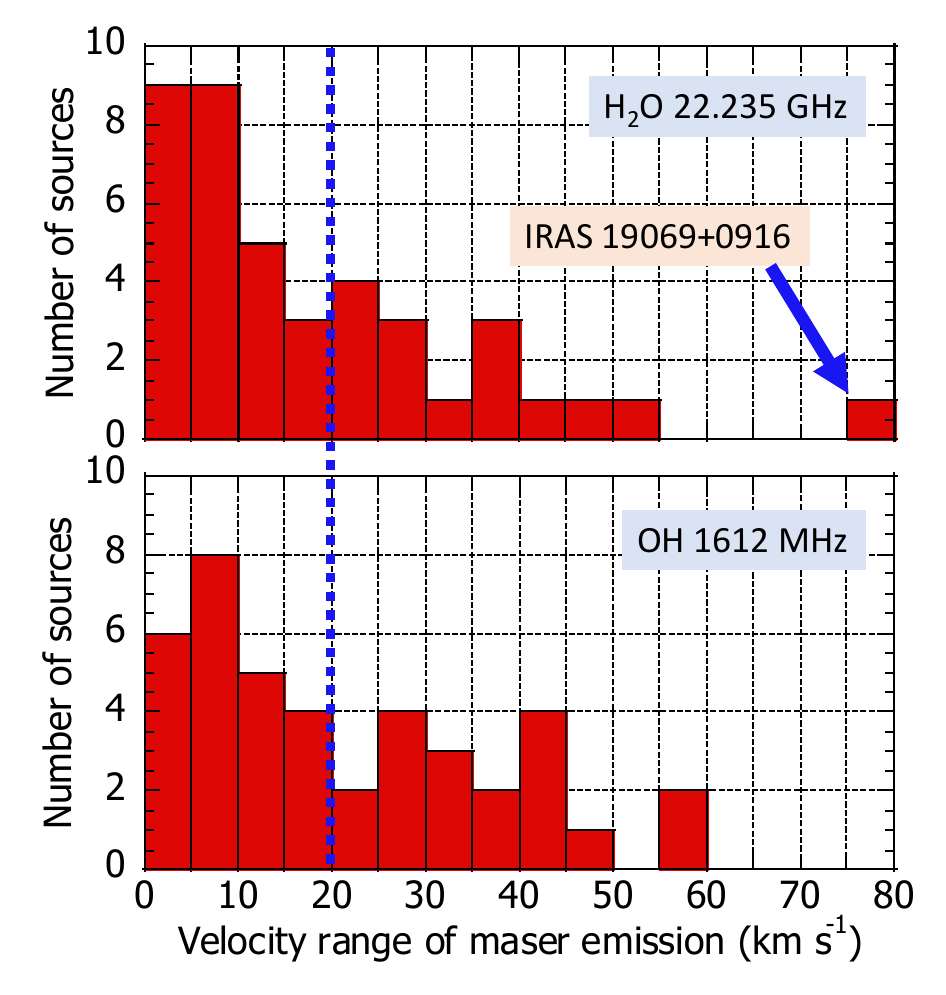}
    \caption{Histograms of the velocity ranges of the H$_2$O 22.235 GHz (upper panel) and OH 1612 MHz (lower panel) maser lines.  Half of the velocity range corresponds to the expansion velocity. The blue vertical dashed line represents a velocity range of 20~km~s$^{-1}$ (corresponding to an expansion velocity of 10~km~s$^{-1}$; see text).}
    \label{fig:vel-range-hist}
\end{figure}

The histograms of the velocity ranges of the OH and H$_2$O maser emission are shown in Figure~\ref{fig:vel-range-hist}. Looking at the lower velocity range below 20~km~s$^{-1}$ (as indicated by the blue vertical dotted line in Figure~\ref{fig:vel-range-hist}), we can see that there are more samples in this range for H$_2$O maser sources than for OH maser sources. As explained in Section~\ref{subsec:basic}, this is due to the fact that the material in the H$_2$O maser emission region is still accelerating, resulting in a slower overall expansion velocity compared to the material in the OH maser emission region. The maximum expansion velocity of the OH maser among all samples is 29.8~km~s$^{-1}$ (the object exhibiting this velocity is IRAS 19083+0851 showing the velocity range of 59.5~km~s$^{-1}$), which is a relatively large value but still possibly within the range of the typical expansion velocity of an AGB envelope (there are previous studies suggesting that IRAS 19083+0851 is possibly an RSG; See, Section \ref{subsec:19083+0851} for details). Notably, IRAS 19069+0916, has a velocity range of H$_2$O maser emission that exceeds this maximum velocity range of OH maser emission. The properties of this object are discussed in detail in Section~\ref{subsec:19069+0916}.

\begin{figure}[h]
    \centering
    \includegraphics[width = 0.9\textwidth]{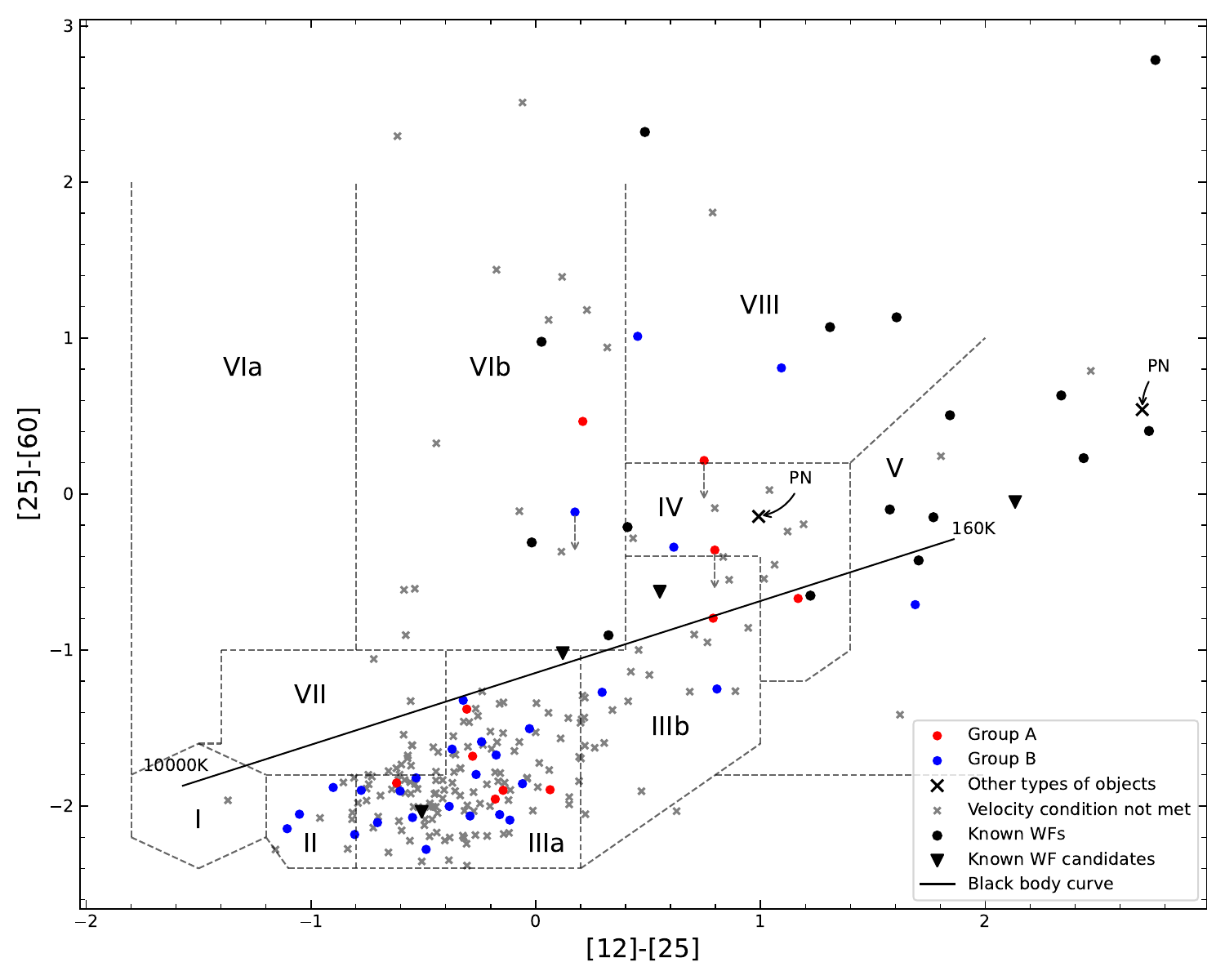}
    \caption{IRAS two-color diagram of 41 sources selected by the velocity condition. This plot uses color indices defined as follows $[12]-[25]=2.5 \log \left(\frac{F_{25 \mu \mathrm{m}}}{F_{12 \mu \mathrm{m}}}\right)$ and $[25]-[60]=2.5 \log \left(\frac{F_{60 \mu \mathrm{m}}}{F_{25 \mu \mathrm{m}}}\right)$. The red and blue dots represent the 41 objects selected by our velocity comparison analysis. ``Group A'' represents sources for which the deviation of the H$_2$O maser velocity range from the OH maser velocity range is reliable; ``Group B'' represents sources for which the velocity deviation is not sufficiently reliable (see, Section~\ref{subsec:indiv} for details). The gray crosses represent the sources that did not meet the velocity condition. The black circles are known WFs, and the black triangles are WF candidates reported by \citet{2013ApJ...769...20Y,2014ApJ...794...81Y}. The black "X" marks represent objects that were included in the 41 selected sources but were clearly not AGB or post-AGB stars. The data points with gray arrows are sources with an IRAS flux quality flag (QF) of 1. The IRAS flux with a QF of 1 is practically an upper limit, and therefore the true value of the IRAS color is assumed to be in the direction of the arrow. The QF values for all sources are summarized in Table~\ref{tab:mira-check}. The black line corresponds to the blackbody radiation curve, and the area separated by the dotted line is the classification of evolved stars given by \citet{1988A&A...194..125V}.}
    \label{fig:2color}
\end{figure}

In Figure~\ref{fig:2color} we have plotted the selected 41 sources on an IRAS two-color diagram to briefly inspect their evolutionary state with respect to mid-infrared colors. The sources are divided into two groups, ``Group A'' (red dots) and ``Group B'' (blue dots), based on the maser profile checks described in Section~\ref{subsec:indiv}  (See also this section for definitions of Group A and B). The areas surrounded by the black dashed lines in the figure indicate the well-known classification of evolved stars by \citet{1988A&A...194..125V}. According to this classification, regions II and IIIa generally correspond to the AGB phase, while regions IIIb, IV, V, and VIII represent the post-AGB and later phases [it should be noted, however, that RSGs often have infrared colors similar to those of AGB/post-AGB stars and cannot be distinguished from AGB/post-AGB stars by their infrared colors alone; See, \citet{2021MNRAS.505.6051J}]. The 41 selected sources are distributed over a very wide range in Figure~\ref{fig:2color}, from the AGB phase to the post-AGB phase and beyond. It is worth noting that quite a large fraction of the 41 selected sources are distributed in regions II and IIIa, corresponding to the AGB phase. In the figure, the black dots and triangles represent known WFs and WF candidates, respectively. Most of the black data points are outside regions II and IIIa, but two sources, IRAS 15193+3132 and IRAS 18056--1514, are within region IIIa. These two sources were previously identified as WF candidates using methods similar to those used in this study \citep{2013ApJ...769...20Y}. The gray crosses in the figure represent the sources for which the velocity range of the H$_2$O maser line did not exceed that of the OH maser line. Of the sources we considered in Group A, IRAS 19319+2214 is in the IIIb region, where early post-AGB stages predominate; IRAS 19069+0916 is located near the boundary between the IIIb and IV regions, which is another source that shows early post-AGB stellar color. Three other Group A sources (IRAS 18251--1048, IRAS 18588+0428, IRAS 19067+0811) are found in the color regions (IV, VIb, VIII), containing more evolved AGB stars
and early  post-AGB stars. These properties will be discussed in the next section.

\begin{deluxetable*}{lrrrrrrl}
\tabletypesize{\scriptsize}
\tablecolumns{8}
\tablecaption{Velocity information for 41 selected sources \label{tab:velocity}}
\tablehead{
\colhead{Source name} & \colhead{$V_b$ (OH)}& \colhead{$V_r$ (OH)} & \colhead{$\Delta V_{\rm OH}$} & \colhead{$V_{\rm min}$ (H$_2$O)} & \colhead{$V_{\rm max}$ (H$_2$O)} & \colhead{$\Delta V_{\rm H_2O}$} & \colhead{Reference for H$_2$O} \\
\colhead{} & \colhead{(km s$^{-1}$)}& \colhead{(km s$^{-1}$)} & \colhead{(km s$^{-1}$)} & \colhead{(km s$^{-1}$)} & \colhead{(km s$^{-1}$)} & \colhead{(km s$^{-1}$)} & \colhead{}
}
\startdata
IRAS 04396+0647 & 15.5 & 20.5 & 5.0 & 12.1 & 14.4 & 2.3 & VAL01 \\
IRAS 05027$-$2158 & $-$15.9 & $-$0.2 & 15.7 & $-$30.5 & $-$25.9 & 4.6 & TAK01 \\
IRAS 05528+2010 & $-$46.0 & $-$39.0 & 7.0 & $-$42.0 & $-$36.0 & 6.0 & NEU21 \\
IRAS 07445$-$2613 & 76.9 & 87.4 & 10.5 & 76.6 & 85.3 & 8.7 & VAL01; KIM13 \\
IRAS 07446$-$3210 & 39.0 & 47.4 & 8.4 & 19.5 & 31.9 & 12.4 & TAK01; KIM10 \\
IRAS 08357$-$1013 & $-$3.9 & 39.4 & 43.3 & 65.0 & 66.6 & 1.6 & TAK01; TAK94 \\
IRAS 13001+0527 & 23.5 & 24.0 & 0.5 & 8.4 & 25.0 & 16.6 & VAL01 \\
IRAS 15060+0947 & $-$16.3 & 1.8 & 18.1 & $-$14.4 & 3.0 & 17.4 & VAL01; TAK01 \\
IRAS 15255+1944 & $-$1.4 & 14.0 & 15.4 & $-$1.8 & 8.8 & 10.6 & VAL01; KIM10 \\
IRAS 15408$-$5413 & $-$105.2 & $-$101.1 & 4.1 & $-$85.7 & $-$75.9 & 9.8 & GOM15A; WAL11 \\
IRAS 17343+1052 & $-$62.6 & $-$53.1 & 9.5 & $-$63.0 & $-$53.0 & 10.0 & YUN13 \\
IRAS 17392$-$3020 & $-$24.9 & 18.5 & 43.4 & $-$18.0 & 21.0 & 39.0 & DEA07 \\
IRAS 17443$-$2949 & $-$16.7 & $-$4.8 & 11.9 & $-$6.4 & 0.9 & 7.3 & GOM08; SUA07 \\
IRAS 17540$-$1919 & 56.4 & 64.0 & 7.6 & 55.0 & 55.0 & 0.0 & CHO17 \\
IRAS 17580$-$3111 & 1.0 & 29.2 & 28.2 & 0.4 & 21.3 & 20.9 & SUA07; GOM08 \\
IRAS 18009$-$2019 & 3.0 & 26.0 & 23.0 & $-$1.5 & 21.0 & 22.5 & VAL01 \\
IRAS 18034$-$2441 & $-$12.7 & 23.7 & 36.4 & $-$13.2 & $-$13.2 & 0.0 & CHO17 \\
IRAS 18050$-$0518 & $-$34.2 & $-$20.3 & 13.9 & $-$32.0 & $-$20.0 & 12.0 & YUN13 \\
IRAS 18052$-$2016 & 23.0 & 80.2 & 57.2 & 14.0 & 43.0 & 29.0 & DEA07 \\
IRAS 18085+0752 & $-$71.2 & $-$63.5 & 7.7 & $-$70.6 & $-$67.9 & 2.7 & VAL01; YUN13; CHO17 \\
IRAS 18251$-$1048 & 71.0 & 111.2 & 40.2 & 70.0 & 110.0 & 40.0 & YUN13 \\
IRAS 18348$-$0526 & 13.2 & 41.7 & 28.5 & 25.8 & 44.7 & 18.9 & TAK01; TAK94 \\
IRAS 18436+4334 & $-$28.2 & $-$13.8 & 14.4 & $-$29.4 & $-$20.2 & 9.2 & TAK01; KIM10 \\
IRAS 18494$-$0130 & 65.8 & 91.7 & 25.9 & 65.0 & 72.0 & 7.0 & YUN13 \\
IRAS 18588+0428 & 29.4 & 74.7 & 45.3 & 29.0 & 74.0 & 45.0 & DEA07 \\
IRAS 19052+0922 & 26.0 & 62.3 & 36.3 & 5.6 & 12.4 & 6.8 & VAL01 \\
IRAS 19067+0811 & 42.3 & 75.9 & 33.6 & 47.2 & 80.0 & 32.8 & VLE14 \\
IRAS 19068+1127 & $-$8.2 & 23.6 & 31.8 & $-$13.0 & 22.0 & 35.0 & YUN13 \\
IRAS 19069+0916 & 10.0 & 54.5 & 44.5 & $-$32.2 & 47.0 & 79.2 & TAK01; VLE14 \\
IRAS 19083+0851 & 16.3 & 75.8 & 59.5 & 7.3 & 60.7 & 53.4 & TAK01; KIM13 \\
IRAS 19229+1708 & 14.6 & 18.7 & 4.1 & 24.3 & 53.0 & 28.7 & VAL01; YUN13 \\
IRAS 19255+2123 & $-$3.0 & 9.0 & 12.0 & 22.4 & 22.6 & 0.2 & GRE04 \\
IRAS 19319+2214 & 12.3 & 29.1 & 16.8 & 8.8 & 33.6 & 24.8 & GOM15A \\
IRAS 19395+1949 & 25.0 & 28.2 & 3.2 & 28.0 & 56.0 & 28.0 & YUN13 \\
IRAS 19422+3506 & $-$65.5 & $-$31.2 & 34.3 & $-$68.5 & $-$33.5 & 35.0 & VLE14 \\
IRAS 19466+2751 & 29.6 & 57.3 & 27.7 & $-$2.8 & 5.0 & 7.8 & VAL01 \\
IRAS 20015+3019 & 5.5 & 7.3 & 1.8 & 16.7 & 20.1 & 3.4 & TAK01 \\
IRAS 20440+0412 & $-$54.8 & $-$49.9 & 4.9 & $-$52.0 & $-$48.0 & 4.0 & YUN13 \\
IRAS 20547+0247 & $-$80.4 & $-$75.3 & 5.1 & $-$75.0 & $-$75.0 & 0.0 & ZUC87 \\
IRAS 22516+0838 & $-$8.6 & 12.3 & 20.9 & $-$7.8 & 14.0 & 21.8 & KIM10 \\
IRAS 22556+5833 & $-$67.0 & $-$59.0 & 8.0 & $-$58.3 & $-$44.8 & 13.5 & VAL01; TAK01 \\
\enddata
\tablecomments{For the OH maser observations, values from Engels \& Bunzel (2015) are used. See Appendix \ref{ap:code} for reference code. }
\end{deluxetable*}

\begin{deluxetable*}{lrrrrrrr}
\tabletypesize{\scriptsize}
\tablecolumns{7}
\tablecaption{Infrared data for 41 selected sources \label{tab:infrared}}
\tablehead{
\colhead{Source name} & \multicolumn{2}{c}{WISE position} & \colhead{Angdist} & \multicolumn{3}{c}{IRAS flux} &  \colhead{QF} \\
\colhead{} & \colhead{R.A.} & \colhead{Dec.} & \colhead{} & \colhead{$F_{\rm 12\mu m}$} & \colhead{$F_{\rm 25\mu m}$} & \colhead{$F_{\rm 60\mu m}$} & \colhead{}  \\
\colhead{} & \multicolumn{2}{c}{(J2000.0)} & \colhead{(arcsec)} & \colhead{(Jy)} & \colhead{(Jy)} & \colhead{(Jy)} & \colhead{}
}
\startdata
IRAS 04396+0647 & 70.5888064  & 6.8778639 & 1.32 & 49.3 & 30.2 & 5.7 & 333 \\
IRAS 05027$-$2158 & 76.2118956  & $-$21.9046758 & 0.36 & 157.0 & 82.1 & 11.8 & 333 \\
IRAS 05528+2010 & 88.9547044  & 20.1751498 & 1.09 & 682.0 & 259.0 & 39.1 & 333 \\
IRAS 07445$-$2613 & 116.6595203  & $-$26.3432689 & 2.71 & 77.9 & 37.1 & 5.0 & 333 \\
IRAS 07446$-$3210 & 116.6422915  & $-$32.3045251 & 0.01 & 117.0 & 93.7 & 21.7 & 333 \\
IRAS 08357$-$1013 & 129.5355866  & $-$10.4042302 & 4.22 & 62.5 & 59.2 & 10.7 & 333 \\
IRAS 13001+0527 & 195.6584898  & 5.1856231 & 9.27 & 462.0 & 226.0 & 39.3 & 333 \\
IRAS 15060+0947 & 227.1072940  & 9.6051642 & 2.73 & 35.2 & 26.9 & 4.0 & 333 \\
IRAS 15255+1944 & 231.9459816  & 19.5643970 & 7.85 & 235.0 & 150.0 & 18.4 & 333 \\
IRAS 15408$-$5413 & 236.1658919  & $-$54.3847874 & 0.25 & 167.0 & 351.0 & 111.0 & 333 \\
IRAS 17343+1052 & 264.1852174  & 10.8518791 & 6.86 & 37.6 & 16.4 & 2.9 & 333 \\
IRAS 17392$-$3020 & 265.6273152  & $-$30.3686911 & 0.87 & 6.3 & 17.3 & 36.4 & 332 \\
IRAS 17443$-$2949 & 266.8971436  & $-$29.8481106 & 3.8 & 15.8 & 39.4 & 34.5 & 333 \\
IRAS 17540$-$1919 & 269.2627897  & $-$19.3377071 & 0.25 & 83.9 & 30.3 & 4.2 & 333 \\
IRAS 17580$-$3111 & 270.3349569  & $-$31.1889743 & 10.28 & 3.2 & 15.3 & 8.0 & 333 \\
IRAS 18009$-$2019 & 270.9855195  & $-$20.3167520 & 3.17 & 419.0 & 294.0 & 46.5 & 333 \\
IRAS 18034$-$2441 & 271.6230128  & $-$24.6799678 & 0.16 & 17.9 & 23.5 & 7.3 & 333 \\
IRAS 18050$-$0518 & 271.9209666  & $-$5.3053559 & 0.32 & 35.7 & 30.8 & 4.6 & 333 \\
IRAS 18052$-$2016 & 272.0678773  & $-$20.2697236 & 1.31 & 21.4 & 32.5 & 82.4 & 333 \\
IRAS 18085+0752 & 272.7440259  & 7.8856846 & 6.41 & 21.7 & 13.1 & 1.9 & 333 \\
IRAS 18251$-$1048 & 276.9846802  & $-$10.7816116 & 4.21 & 6.7 & 13.3 & 16.2 & 331 \\
IRAS 18348$-$0526 & 279.3854407  & $-$5.3997526 & 0.09 & 360.0 & 634.0 & 463.0 & 333 \\
IRAS 18436+4334 & 281.2922769  & 43.6354216 & 4.44 & 64.6 & 50.6 & 9.7 & 333 \\
IRAS 18494$-$0130 & 283.0060527  & $-$1.4461850 & 0.13 & 12.6 & 14.8 & 13.3 & 331 \\
IRAS 18588+0428 & 285.3338523  & 4.5420510 & 1.07 & 10.8 & 13.1 & 20.1 & 332 \\
IRAS 19052+0922 & 286.9128078  & 9.4543756 & 3.06 & 24.8 & 18.7 & 5.3 & 333 \\
IRAS 19067+0811 & 287.2846276  & 8.2761164 & 0.22 & 24.6 & 72.1 & 38.9 & 333 \\
IRAS 19068+1127 & 287.2986926  & 11.5466987 & 12.02 & 18.1 & 19.2 & 3.4 & 333 \\
IRAS 19069+0916 & 287.3302736  & 9.3532127 & 2.4 & 4.8 & 10.0 & 7.2 & 331 \\
IRAS 19083+0851 & 287.6972070  & 8.9396218 & 1.22 & 41.6 & 36.4 & 6.3 & 332 \\
IRAS 19229+1708 & 291.3021044  & 17.2469940 & 5.71 & 46.1 & 35.6 & 7.6 & 333 \\
IRAS 19255+2123 & 291.9334338  & 21.5009583 & 0.05 & 2.4 & 29.3 & 48.2 & 333 \\
IRAS 19319+2214 & 293.5149397  & 22.3544324 & 3.28 & 3.2 & 6.6 & 3.2 & 333 \\
IRAS 19356+0754 & 294.5049011  & 8.0257915 & 0.21 & 1.1 & 8.0 & 7.6 & 333 \\
IRAS 19422+3506 & 296.0292220  & 35.2355954 & 0.14 & 203.0 & 172.0 & 28.4 & 333 \\
IRAS 19466+2751 & 297.1552342  & 27.9956429 & 15.74 & 27.8 & 27.1 & 6.8 & 333 \\
IRAS 20015+3019 & 300.9103669  & 30.4692632 & 0.05 & 162.0 & 115.0 & 25.5 & 333 \\
IRAS 20440+0412 & 311.6382059  & 4.3934590 & 1.61 & 9.1 & 5.2 & 0.9 & 333 \\
IRAS 20547+0247 & 314.3179038  & 2.9791841 & 1.14 & 45.5 & 33.8 & 10.0 & 333 \\
IRAS 22516+0838 & 343.5474973  & 8.9019877 & 0.37 & 113.0 & 63.9 & 11.6 & 333 \\
IRAS 22556+5833 & 344.4207333  & 58.8201558 & 7.17 & 189.0 & 170.0 & 24.8 & 333 \\
\enddata
\tablecomments{QF: The quality flag for the IRAS flux density, where 3 means high quality, 2 moderate quality, and 1 an upper limit. The numbers correspond to 12, 25, and 60 micron flux densities from left to right. Angdist: Angular separation between the location of the maser sources in the database and that of the AllWISE counterpart.}
\end{deluxetable*}

\subsection{Characteristics of individual sources} \label{subsec:indiv}

At the beginning of Section~\ref {subsec:overview}, we mentioned that 17.9\% of the sources for which velocity comparisons were made met the velocity requirements as WF candidates. This rate value seems too high for a percentage of WFs, and the reliability of the maser data used and the contamination in the sample must be carefully considered.
For this purpose, here, we discuss the maser characteristics of individual objects using available information from the literature. Specifically, we consider the following perspectives:

\begin{enumerate}

\item Whether the OH maser line profiles indeed show double-peaked profile typically seen towards AGB stars . A considerable number of sources classified as such profile sources have a small velocity gap between the two peaks (a few~km~s$^{-1}$ or less). The line profiles are informative to consider whether such short-spaced double peaks is really associated with, respectively, the receding and approaching sides of a spherically expanding CSE. We also checked for any velocity components that were slipped from the database.

\item Whether there are any velocity components of the H$_2$O maser line that we have missed. In particular, the weak velocity component at velocities beyond the OH maser velocity range was carefully inspected. We also checked whether the H$_2$O maser line profiles were similar to known WFs. The known WFs share the common features of the H$_2$O maser spectra as follows: (1) the velocity range is larger than 100 km~s$^{-1}$, and (2) two or more peaks are always or often found. The intensity peaks appear randomly at different radial velocities and there is no fixed pattern: Sometimes they exhibit  double peaks or multiple peaks. However, the H$_2$O maser spectra used for confirmation were obtained by a variety of different instruments, and the S/N ratios achieved vary from observation to observation. It should be added, therefore, that not all spectra meet the S/N ratios necessary to distinguish the above conditions. For the reader's reference, the papers giving the spectra of  H$_2$O and OH maser lines of known WFs are summarized in Appendix~\ref{ap:known WF masers}.

\item Comparison of SiO and OH maser velocities. Since SiO maser lines are expected to be close to the systemic velocity of the source, the comparison with the OH maser provides a guideline for considering whether the OH maser peak is associated with the approaching or receding side. In addition, a large SiO maser linewidth (e.g. $>$20~km~s$^{-1}$) is a hint for a possible RSG.

\item Comparison of line profiles between the OH 1612 and 1665/1667 MHz lines. Since the 1665 and 1667 MHz lines have pumping route different from that of  the 1612 MHz line, the difference of the OH line profiles provides a clue to the nature of CSE. We also checked the intensity ratio between the 1612 MHz and the 1665/1667 MHz lines for sources where all three lines are detected. Typically, the 1612 MHz line is weaker than the 1665 and 1667 MHz lines in YSOs \citep[see, e.g.,][]{1996A&ARv...7...97H}. In fact, there exist no sources showing the weaker 1612 MHz intensity than the 1665/1667 MHz lines.

\item We also checked the literature information available from SIMBAD to see if it had been identified in objects other than AGB and post-AGB stars. 

\end{enumerate}

The details of the above confirmations are summarized for each source in the subsections below and in Appendix~\ref{ap:indiv} and \ref{ap:other}. The literature search with SIMBAD database revealed that two of the 41 selected objects (IRAS 17443$-$2949 and IRAS 19255+2123) are identified as planetary nebulae. The remaining 39 sources are classified into ``Group A'' and ``Group B'' based on the results of the above confirmation.  Group A includes sources that meet either of the following conditions: (1) it is fairly certain that the velocity range of the H$_2$O maser exceeds that of the OH maser, and the H$_2$O maser line profile resembles known WFs, or (2) even if condition (1) is not met, judging from the maser characteristics found in the literature, it is unlikely to be an AGB or post-AGB star with a normal spherically expanding circumstellar envelope.

Group B, on the other hand, includes objects where velocity range excess of the H$_2$O maser is uncertain and the difference in the maser line profile from those of normal AGB or post-AGB stars is unclear. 

To avoid making the main text too long, the following subsections describe only the sources in Group A. Descriptions for the sources in Group B and those identified as non-AGB/post-AGB stars are given in the Appendix~\ref{ap:indiv} and \ref{ap:other}.  For Group A sources, we note in the description of each object whether each object meets condition (1) or (2) above.

\begin{deluxetable*}{lllllll}
\tabletypesize{\scriptsize}
\tablecolumns{6}
\tablecaption{Variable star names and source classification \label{tab:mira-check}}
\tablehead{
\colhead{Source name} & \colhead{Group}& \colhead{Region} & \colhead{Variable star name} & \colhead{JE21} & \colhead{AAVSO} & \colhead{Note}
}
\startdata
IRAS 04396+0647 & B & IIIa & BZ Tau & 1 & --- &  \\
IRAS 05027$-$2158 & B & IIIa & T Lep & --- & mira-like &  \\
IRAS 05528+2010 & B & II & U Ori & --- & mira-like &  \\
IRAS 07445$-$2613 & B & II & SS Pup & --- & mira-like &  \\
IRAS 07446$-$3210 & B & IIIa & V353 Pup & --- & --- &  \\
IRAS 08357$-$1013 & B & IIIa & NSV 17942 & --- & --- &  \\
IRAS 13001+0527 & B & IIIa & RT Vir & --- & mira-like &  \\
IRAS 15060+0947 & B & IIIa & FV Boo & 1 & mira-like &  \\
IRAS 15255+1944 & B & IIIa & WX Ser & --- & mira-like &  \\
IRAS 15408$-$5413 & B & IIIb & --- & --- & --- &  \\
IRAS 17343+1052 & B & II & V790 Oph & --- & --- &  \\
IRAS 17392$-$3020 & B & VIII & --- & --- & --- &  \\
IRAS 17443$-$2949 & --- & IV & --- & --- & --- & Known PN \\
IRAS 17540$-$1919 & B & II & VV Sgr & --- & --- &  \\
IRAS 17580$-$3111 & B & V & --- & --- & --- &  \\
IRAS 18009$-$2019 & B & IIIa & V4120 Sgr & --- & mira-like &  \\
IRAS 18034$-$2441 & B & IIIb & --- & --- & --- &  \\
IRAS 18050$-$0518 & B & IIIa & --- & --- & --- &  \\
IRAS 18052$-$2016 & B & VIII & --- & --- & --- &  \\
IRAS 18085+0752 & B & IIIa & XY Oph & 1 & --- &  \\
IRAS 18251$-$1048 & A & VIII & --- & --- & --- &  \\
IRAS 18348$-$0526 & B & IV & --- & --- & --- &  \\
IRAS 18436+4334 & B & IIIa & RW Lyr & --- & mira-like &  \\
IRAS 18494$-$0130 & B & VIb & --- & --- & --- &  \\
IRAS 18588+0428 & A & VIb & --- & --- & --- &  \\
IRAS 19052+0922 & A & IIIa & --- & 1 & --- &  \\
IRAS 19067+0811 & A & IV & --- & 1 & --- &  \\
IRAS 19068+1127 & A & IIIa & --- & 1 & --- &  \\
IRAS 19069+0916 & A & IV & --- & 1 & --- &  \\
IRAS 19083+0851 & A & IIIa & --- & 4 & --- &  \\
IRAS 19229+1708 & A & IIIa & --- & 4 & --- & Gaia DR3 suggests LPV \\
IRAS 19255+2123 & --- & V & --- & --- & --- & Known PN \\
IRAS 19319+2214 & A & IIIb & --- & 2 & --- &  \\
IRAS 19395+1949 & B & IIIa & --- & 1 & --- &  \\
IRAS 19422+3506 & A & IIIa & --- & 1 & --- &  \\
IRAS 19466+2751 & B & IIIa & --- & --- & --- &  \\
IRAS 20015+3019 & B & IIIa & V719 Cyg & 4 & --- &  \\
IRAS 20440+0412 & B & IIIa & BR Del  & 1 & mira-like &  \\
IRAS 20547+0247 & B & IIIa & U Equ & 2 & --- &  \\
IRAS 22516+0838 & A & IIIa & KZ Peg & 1 & --- &  \\
IRAS 22556+5833 & B & IIIa & --- & --- & --- &  \\
\enddata
\tablecomments{TE21: ``Group" in Table~1 of Jimenez-Esteban et al. (2021).  1 = LPLAV, 2 = Post-AGBs, 3 = Unclassified, 4 = YSO; RSG; False OH detection; Irregular; Optical low-amplitude regular variability.\\
AAVSO: Results of checking the AAVSO light curves. The souces that are considered to show Mira-type pulsation are marked as "mira-like".}
\end{deluxetable*}


\subsubsection{IRAS 18251--1048} \label{subsec:18251-1048}
The profile of the OH 1612 MHz line is typically double-peaked, with some fine structure observed in the red-shifted component, as reported in previous studies by \citet{1989A&AS...78..399T} and \citet{2014ApJ...794...81Y}. The OH 1667 MHz line is weakly detected, with peaks located near 115~km~s$^{-1}$ and 20~km~s$^{-1}$, as reported in \citet{2014ApJ...794...81Y}. The H$_2$O maser line profile is double-peaked, according to observations by \citet{2013ApJ...769...20Y} and \citet{2017ApJS..232...13C}, and the emission is mostly within the velocity range of the OH maser line. The information in the database suggests that the deviation of the H$_2$O maser line from the OH maser velocity range is 1.0~km~s$^{-1}$. The SiO $v=1$ and 2, $J=1-0$ lines have been detected  near 90~km~s$^{-1}$, and the line profiles are roughly single-peaked, as reported in previous studies by \citet{2003PASJ...55..229N} and \citet{2017ApJS..232...13C}. 
Based on the comparison of the OH and H$_2$O velocity ranges, it is difficult to conclude that the H$_2$O maser velocity deviations are real.
However, it is worth noting that the OH 1667 MHz line deviates from the velocity range of the OH 1612 MHz line (see, Section~\ref{subsec:velocity}). This source satisfies condition (2).

\subsubsection{IRAS 18588+0428} \label{subsec:18588+0428}
The profile of the OH maser line shows a typical double-peak profile \citep{1994ApJS...93..549L,2004ApJS..155..595D}. According to the spectra in the literature, the OH maser emission appears to be detected even outside the redshifted peak, near 20~km~s$^{-1}$. The H$_2$O maser spectrum shows many peaks, which are uniformly distributed within the velocity range of the OH maser line \citep{2007ApJ...658.1096D}. The SiO $v=1$ and 2, $J=1-0$ lines are detected  near 55~km~s$^{-1}$, and the line profiles are a typical single peak \citep{2004PASJ...56..765D,2014ApJS..211...15Y}. The deviation of the H$_2$O maser line from the velocity range of the OH maser is small (0.4~km~s$^{-1}$). However, it is worth noting that the H$_2$O maser profile with many peaks is similar to the known WF, IRAS 18286$-$0959 \citep{2011ApJ...741...94Y}, except for the H$_2$O velocity range not largely exceeding that of the OH maser line. This source satisfies condition (2).

\subsubsection{IRAS 19052+0922} \label{subsec:19052+0922}
The OH maser shows a typical double-peak profile, as reported in the literature such as \citet{1994ApJS...93..549L}. For this source there are two H$_2$O maser observations in the literature. One by \citet{2001A&A...368..845V} and another by \citet{1996A&AS..116..117E}. \citet{1996A&AS..116..117E} observed a double peaked profile at 37.8 and 54.8~km~s$^{-1}$, which fits well with the OH maser velocity. The H$_2$O maser profile in \citet{2001A&A...368..845V} has a single peak at 10~km~s$^{-1}$. The 10~km~s$^{-1}$ peak of \citet{2001A&A...368..845V} was apparently not present in the spectra of \citet{1996A&AS..116..117E}. The 10~km~s$^{-1}$ peak is clearly outside the velocity range of the OH maser line (the peak velocity of the OH red-shifted component is near 26~km~s$^{-1}$). The SiO $v=1$, $J=1-0$ line is detected, as reported in \citet{2017ApJS..232...13C}, with a typical single peak profile and peak velocity near 40~km~s$^{-1}$. Although the H$_2$O maser profile is single-peaked, it should be considered a WF candidate according to our criteria. This source satisfies condition (1).

\subsubsection{IRAS 19067+0811} \label{subsec:19067+0811}
The OH maser profile shows a typical double-peak profile, as reported in references \citep{1989A&AS...78..399T,1994ApJS...93..549L, 2012A&A...537A...5W}. The H$_2$O maser profile is also double-peaked, with some of the fine components on the red-shifted side detected outside the velocity range of the OH maser line, as reported in \citet{2014A&A...569A..92V}. The SiO $v=1$, 2 and 3, $J=1-0$ lines, SiO $v=1$, $J=2-1$ lines, and $^{29}$SiO $v=0$, $J=1-0$ and $2-1$ lines have been detected, as reported in \citet{2007ApJ...669..446N}. All the SiO line profiles are single-peaked with a peak velocity near 64~km~s$^{-1}$. The deviation of the velocity of the H$_2$O maser line from the OH maser velocity range is 4.1~km~s$^{-1}$ according to the database, but there may be even faster velocity components when looking at the spectra in the literature. The H$_2$O maser line exhibits a double-peaked profile reminiscent of a bipolar flow. \citet{2014A&A...569A..92V} made mapping observations in H$_2$O maser line with JVLA, which did not show a narrowly collimated jet. However, the overall velocity field can be interpreted as a loosely squeezed bipolar flow. This source satisfies condition (2).

\subsubsection{IRAS 19068+1127} \label{subsec:19068+1127}

The OH maser line shows a typical double-peak profile, as reported in the literature such as \citet{1994ApJS...93..549L}. The H$_2$O maser profile has multiple peaks and is distributed throughout the velocity range of the OH maser line, as reported in the literature such as \citet{2013ApJ...769...20Y}. This source is similar to IRAS 18286--0959 \citep{2011ApJ...741...94Y}, except that the velocity range of the H$_2$O maser line does not largely exceed the OH maser's velocity range. The minimum velocity of the H$_2$O maser line deviates from the OH maser's velocity range by 4.8~km~s$^{-1}$, according to the database. The H$_2$O spectrum from \citet{2013ApJ...769...20Y} does not clearly show the velocity excess suggested by the database. This source satisfies condition (2).

\subsubsection{IRAS 19069+0916} \label{subsec:19069+0916}
The OH maser profile shows a typical double-peak profile  \citep{1988ApJS...66..183E,1994ApJS...93..549L}. Three H$_2$O maser spectra were found in the literature \citep{2001PASJ...53..517T, 2014A&A...569A..92V,2010ApJS..188..209K}, but the line profiles show very large temporal variability. Looking at all the spectra in the literature, emission lines appear over velocities from near $-$35~km~s$^{-1}$ to near 45~km~s$^{-1}$. The velocity range of this H$_2$O maser completely deviates from that of the OH maser line. The $-$35~km~s$^{-1}$ component was detected only by \citet{2001PASJ...53..517T}. Mapping observations in the H$_2$O maser line have been made by \citet{2014A&A...569A..92V} using JVLA, but the $-$35~km~s$^{-1}$ component was not detected. The SiO $v=1$ and 2, $J=1-0$ lines are detected near 30~km~s$^{-1}$, and the line profiles are typical single peaks \citep{2003PASJ...55..229N,2010ApJS..188..209K}. This source satisfies condition (1).

\subsubsection{IRAS 19083+0851} \label{subsec:19083+0851}
The OH maser spectrum in \citet{1993ApJS...89..189C} show three peaks at around 17~km~s$^{-1}$, 37~km~s$^{-1}$, and 75~km~s$^{-1}$. Two H$_2$O maser spectra were found in the literature \citep{1996A&AS..116..117E, 2013AJ....145...22K}, showing strong time variation in its profile. The H$_2$O maser profile is characterized by multiple peaks over a wide velocity range, similar to the profile of the known WF, IRAS 18286$-$0959 \citep{2011ApJ...741...94Y}. The minimum velocity of 7.3~km~s$^{-1}$ given in the database was not confirmed by the H$_2$O spectra in the literature. But, in the spectrum given by \citet{1996A&AS..116..117E}, a weak H$_2$O maser emission is detected around 27~km~s$^{-1}$, implying the existence of high-velocity components on the blue-shifted side. The SiO $v=1$ and 2, $J=1-0$ lines have been detected \citep[ near 56~km~s$^{-1}$,][]{2013AJ....145...22K}. The $v=1$ line shows five peaks, and the $v=2$ line shows double peaks, with line widths as wide as 15--20~km~s$^{-1}$ at zero-intensity levels. This differs from the single-peak profile seen in a spherically expanding envelope of an AGB star. In any case, the information in the database supports the deviation of the H$_2$O maser velocity from the range of the OH maser velocity. This source is suggested to be an RSG because of the relatively large velocity range of the OH maser \citep{1996A&AS..116..117E} and the small amplitude of the light curves in the near-infrared bands \citep{2021MNRAS.505.6051J}. However, similar properties may also be found in early post-AGB stars.  This source satisfies condition (1).

\subsubsection{IRAS 19229+1708} \label{subsec:19229+1708}
The OH maser line has a typical double-peaked profile, with peak velocities at 14.6~km~s$^{-1}$ and 18.7~km~s$^{-1}$. However, a spectrum presented in \citet{1990ApJ...362..634L} shows a sharp peak near 15~km~s$^{-1}$, with a decreasing intensity on the blue-shifted side and a slower decrease on the red-shifted side. This profile is reminiscent of the blue-shifted component of the double-peak profile of the OH maser line emitted from a typical spherically expanding envelope of an AGB star. There are two H$_2$O maser spectra in the literature \citep{2001A&A...368..845V, 2013ApJ...769...20Y} showing multiple peaks within a velocity range of about 50~km~s$^{-1}$, with strong time variability. The most red-shifted peak is detected near 51~km~s$^{-1}$.  The 51~km~s$^{-1}$ component of the H$_2$O maser line is 36.4~km~s$^{-1}$ away from the 14.6~km~s$^{-1}$ peak of the OH maser line. Assuming the approaching side of the spherically expanding envelope is detected in the OH maser line, and the 51~km~s$^{-1}$ component of the H$_2$O maser line corresponds to the receding side, the expansion velocity of the envelope is calculated to be 18.2~km~s$^{-1}$.  This source satisfies condition (2).

\subsubsection{IRAS 19319+2214} \label{subsec:19319+2214}
The OH maser profile is characterized by typical double peaks, but there also exist wing components outside of the two peaks, as seen in the spectrum from \citet{1990ApJ...362..634L}. The database indicates that the velocity range of the H$_2$O maser deviates from that of the OH maser on both the red- and blue-shifted sides. However, the spectra in the literature only show deviation on the red-shifted side. The H$_2$O maser profile also shows multiple peaks, similar to the known WF IRAS 18286$-$0959 \citep{2011ApJ...741...94Y}.  This source satisfies condition (1).

\subsubsection{IRAS 19422+3506} \label{subsec:19422+3506}
The OH maser profile is a double peak, which is typically emitted from a spherically expanding envelope of an AGB star, as reported in \citet{1992A&A...254..133L}. The H$_2$O maser line, however, is detected from a wide velocity range that extends beyond the velocity range of the OH maser line, and its profile varies significantly depending on the time of observation, as reported in \citet{2014A&A...569A..92V}. The H$_2$O maser line of this source has been mapped using the JVLA \citep{2014A&A...569A..92V}, and the resulting map clearly shows structures that are not spherically symmetric, but it is not a collimated bipolar flow.  This source satisfies condition (1).

\subsubsection{IRAS 22516+0838} \label{subsec:22516+0838}
The OH maser profile is basically double-peaked, but it is a rather complex profile with multiple peaks on both the red-shifted and blue-shifted sides \citep{1988ApJS...66..183E,2012A&A...537A...5W}. The OH 1667 MHz line is also detected at a velocity that deviates from the velocity range of the OH 1612 MHz line, which suggests that there is fast molecular gas motion inside the OH 1612 MHz line emission region \citep{1979AJ.....84..345S}. The H$_2$O maser profile is double-peaked and deviates from the OH maser velocity range by about 1.7~km~s$^{-1}$ \citep{2010ApJS..188..209K}. The SiO $v=1$ and 2, $J=1-0$ lines have also been detected \citep{2007ApJ...669..446N}. The SiO maser profile shows a strong peak around the center in the OH maser velocity range, but also broad, weak fast-wing components are seen in the spectra. Although the deviation of the H$_2$O maser velocity from the OH maser velocity range is small, the profiles of the OH main lines and SiO maser line suggest that there are kinematic components inside the envelope that deviate from the spherically expanding motion. This source satisfies condition (2).

\section{Discussions} \label{sec:Discussions}

\subsection{Problems related to making maser source pairs for velocity comparison} \label{subsec:pairs}

In our analysis we compared the velocity ranges of the H$_2$O and OH maser lines for 229 sources. This looks like a relatively small number, considering that there are 2,195 OH (1612 MHz) maser sources and 3,642 H$_2$O (22.235 GHz) maser sources in the databases. Only 10.4\% of the total number of OH maser sources and 6.2\% of the total number of H$_2$O maser sources could be paired for velocity comparison. Of course, it is known that the molecular species of the detected maser lines differ to some extent depending on the evolutionary stage of the evolved star \citep[i.e., Lewis' chronology; see, e.g.,][]{1989ApJ...338..234L,1996A&ARv...7...97H}. However, such an explanation alone cannot account for the small overlap between OH and H$_2$O maser detections.

The low rate of matched pairs compared to the total number of maser sources is due to several reasons. First, for many OH maser sources discovered by blind surveys, follow-up observations to detect H$_2$O masers are not available.  Second, the number of stars with OH masers that also exhibit H$_2$O maser emission is only about 50\% \citep{1996A&AS..116..117E}.  Conversely, not all stars with H$_2$O maser emission have OH masers, especially stars with blue IRAS [12]--[25] colors, where the OH masers are too weak, if present at all, to be detected \citep{1995MNRAS.274..439L}. Finally, pairing the velocities of OH and H$_2$O maser sources requires reasonably accurate coordinates for a match with an AllWISE and/or 2MASS counterpart. Most of the observations in the database were made with single-dish telescopes, resulting in low angular resolution of up to 1$'$ in some cases. In addition, many survey observations of maser sources in the 1990s and 2000s used coordinates from the IRAS Point Source Catalog (PSC), which are known to have uncertainties over 30$''$ in some cases, as noted in studies such as \citet{1998PASJ...50..597D}.

Another point we should note is that in this study we have only selected maser sources that have a clear near-infrared point source as their counterpart. In the future, we will need to confirm the presence of WF jets with a high-resolution radio interferometer, such as VLBI, to confirm whether our selected sources have WF status. To do this, we need precise coordinate information (at least within a few arcseconds). Therefore, as explained in Section~\ref{subsec:h2o-catalog}, we screened out those sources that had multiple possible counterparts within 20$''$--30$''$ and could not be narrowed down to a single counterpart. As the angular resolution of OH and H$_2$O maser observations improves in the future through increased opportunities for radio interferometric observations, the problems described here may disappear.

\subsection{Velocity comparison OH and H$_2$O maser lines and their profiles} \label{subsec:velocity}

The first point to consider is the classification of the line profiles given in the OH maser source database. Due to limitations on the amount of information that can be included as text in the database, it is obviously not possible to provide extensive details on the line profiles.

As mentioned in Section~\ref{subsec:oh-catalog}, each OH maser source in the database is classified as either double peaked (D) or single peaked (S). When we examined the spectra in the literature, we found that sources classified as ``S'' have simple, single peaks in fact. The ``D" sources, on the other hand, exhibit a wide variety of line profiles. As stated in Section~\ref{subsec:basic}, for the purpose of finding WF candidates, we expect double peaks to be caused by the receding and approaching sides of a spherically expanding envelope. However, upon further examination of the literature, we found that a significant number of OH maser sources classified as ``D" do not fit this expectation.

When the velocity interval between the two peaks is as small as a few km~s$^{-1}$, both peaks may be associated with either side of a spherically expanding envelope, since such a small interval cannot be explained as the expansion velocity of an AGB envelope. 

The line profile is very helpful in assessing such a situation. In the case of a typical double peak profile associated with the receding and approaching sides of a spherically expanding envelope, the side closer to the systemic velocity of each peak will have a gradual decrease in intensity, while the far side will be steeply truncated. With a good signal-to-noise ratio, it may be possible to distinguish between these cases by looking at the spectrum alone. However, if the signal-to-noise ratio is not high enough, it may be difficult to tell from the OH maser spectrum alone whether we are looking at a spherically expanding envelope with a slow expansion velocity or the receding or approaching side of a spherically expanding envelope. Even in this case, we have found that a comparison between the OH and SiO maser lines is useful to consider the situation. SiO maser emission lines are often detected near the systemic velocity of the CSEs \citep[see, e.g.,][]{1991A&A...242..211J}. Thus, for example, if both OH double peaks lie to one side of the SiO peak velocity, we can assume that both are emitted from either the receding or the approaching side of a spherically expanding envelope.

Another point to note here is that there may be a significant number of velocity components in the spectra that are missing from the database records. From our inspection of spectra in the literature, there are two cases where they are missing. One is when the intensity of the emission lines is so weak that they are missed. The other is a time variation of the maser line profile. In fact, when examining the spectra, faint features are often found outside the velocity range given in the database. We have also found cases where the line profiles vary considerably with time (e.g. IRAS 19083+0851, IRAS 19229+1708, IRAS 19422+3506, etc.). For sources with large variations, there is naturally a greater chance of missing high velocity components outside the OH velocity range. The profiles of OH and H$_2$O maser lines emitted from the CSEs of evolved stars can vary on timescales of about a month \citep[see, e.g.,][]{2000A&AS..146..179E,2002A&A...388..252E,2003A&A...403L..51E,2016JKAS...49..261K}. Therefore, to fully resolve this problem, it would be necessary to perform monitoring observations for all sources with time intervals of several months, leaving aside the question of whether this can actually be done with existing instruments.

In addition, we note the potential usefulness of the OH maser main lines (i.e., the OH 1665 MHz and 1667 MHz lines). Since the pumping route of the main lines are different from the satellite line, the main lines and satellite lines are thought to trace parts of the circumstellar envelope in different physical conditions. Therefore, comparing the OH 1612 MHz and OH 1665/1667 MHz lines provides similar information to comparing the OH 1612 MHz line and the H$_2$O maser line. For example, as discussed in Section~\ref{subsec:18251-1048}, IRAS 18251-1048 has a velocity range for the 1667 MHz line that exceeds the velocity range for the 1612 MHz line. The OH 1612 MHz and 1665/1667 MHz lines can usually be observed simultaneously with a single instrument, which requires less observing effort than comparing the OH and H$_2$O maser lines. As more radio interferometers become available to observe the OH maser lines in the future, this comparison may be a useful tool for finding evolved stars that are out of spherical symmetry.

The issues discussed here with the OH maser database show the limitations and emphasize the need to evaluate the original maser spectra to improve the velocity ranges obtained from the database. Due to the chosen database format, which gives the velocities of the two strongest maser peaks (for masers classified as "D"), the full complexity of the spectra is not adequately reproduced and the OH maser velocity range is sometimes underestimated.

\subsection{Confirmation of the methodology using known WFs} \label{subsec:confirmation}

We have tested what happens when known WFs are run through our screening process to consider what conditions affect the completeness of the screening. However, in this discussion we will not consider the classical WF condition that the velocity range of the H$_2$O maser emission exceeds 100 km~s$^{-1}$. Since the velocity range of the H$_2$O maser is always considered first, known WFs will never be missed if this condition on the H$_2$O velocity range is included. In the following discussion, we will only consider the comparison of the H$_2$O maser velocity range with the OH maser, as applied in the present work. In this case, the results of the screening showed that only 7 out of 16 known WFs passed. The passed objects include IRAS 15544--5332, IRAS 16342--3814, IRAS 18043--2116, IRAS 18286--0959, IRAS 18450--0148, IRAS 18455+0448, and IRAS 18596+0315.

The 9 known WFs did not pass our screening for the following three reasons. The first reason is that the corresponding OH maser source is not found (i.e., the 1612 MHz OH maser line has been observed in the past but not detected), and three WFs fall into this case: i.e. IRAS 16552--3050, IRAS 17291--2147, and IRAS 19134+2131. 

The second reason is that the velocity range in the database does not meet the WF requirement. IRAS 18460--0151 did not pass screening for this reason. As mentioned in Section~\ref{subsec:velocity}, the velocity range information in the database is not always complete due to time variations in the profiles and other reasons.  It is clear that inspection of line profiles, as we did in Section~\ref {subsec:indiv}, are important to reduce this problem.

The third reason was that no infrared counterpart was found, or we could not converge the number of infrared counterparts into one object. The following five objects fit this reason: IRAS 15103--5754, IRAS 15445--5449, IRAS 18113--2503, IRAS 18139--1816, and IRAS 19190+1102. As discussed in Section~\ref{subsec:velocity}, this problem is primarily due to the listing of poor coordinates in the databases, which were taken from the papers describing the discovery observations.

The above results suggest that to improve the completeness of our screening method, (1) the H$_2$O and OH maser observations must be fully achieved for all sources, (2) multiple observations must be made for a single source to deal with temporal variations in maser intensity and profile, (3) radio observations should provide accurate source positions for comparison with infrared data.

\subsection{Distribution of selected sources on the IRAS two-color diagram} \label{subsec:IIIb}

Many previous studies have often assumed that the structure of the CSE begins to deviate from spherical symmetry in the post-AGB phase. In Figure~\ref {fig:2color}, the IIIb region adjacent to the IIIa region (AGB region) appears to correspond to the early post-AGB region. After careful checking of the spectra, as mentioned in Section~\ref{subsec:indiv}, one source (i.e., IRAS 19319+2214) remained in this region that is certain to exhibit the velocity deviation of the H$_2$O maser line. Given the uncertainty of the IRAS 60 $\mu$m flux, we can assume that another source IRAS 19069+0916 is also substantially in the IIIb region. No radio interferometric observations have been made for IRAS 19319+2214 in the past. We note that the profile of the H$_2$O maser line of this source is similar to that of known WFs (see, Section~\ref{subsec:19319+2214}). The H$_2$O maser line of IRAS 19069+0916 was observed with the VLA \citep{2014A&A...569A..92V}. However, due to the time variation of the line profile, the timing of the VLA observation was not appropriate to confirm the WF status (see, Section~\ref{subsec:19069+0916} for details).

The II and IIIa regions in Figure~\ref{fig:2color} contain mainly (oxygen-rich) AGB stars. As mentioned above, since previous studies have often assumed that the structure of the CSE begins to deviate from spherical symmetry in the post-AGB phase, the identification of WF candidates in these color regions must be carefully considered. Our screening selected 27 sources in the II and IIIa regions. After carefully inspecting the spectra, no sources remained in Region II that showed a clear velocity deviation of the H$_2$O maser line. In Region IIIa, on the other hand, six sources remained with a clear deviation of the H$_2$O maser emission. The six sources are IRAS 19052+0922, IRAS 19068+1127, IRAS 19083+0851, IRAS 19229+1708, IRAS 19422+3506, and IRAS 22516+0838. We would first need to verify if these six sources are AGB stars.

One of the observational characteristics of AGB stars is that many of them are pulsating variables (i.e., Mira-type variables). To confirm whether the star is pulsating, we checked the AAVSO light curves and the light curves of the visible-near-infrared monitoring observations of the Arecibo samples by \citet{2021MNRAS.505.6051J}: See, Table~\ref{tab:mira-check}. First, the four sources IRAS 19052+0922, IRAS 19068+1127, IRAS 19422+3506, and IRAS 22516+0838 have been confirmed as LPLAVs (long-period and large-amplitude variables) in \citet{2021MNRAS.505.6051J} and are considered AGB stars. For IRAS 19422+3506, a mapping observation in the 22.235 GHz H$_2$O maser line has been made with JVLA \citep{2014A&A...569A..92V}, and the non-spherically symmetric structure of the envelope has been confirmed. IRAS 19422+3506 is a pulsating AGB star, but the fact that the CSE has already deviated from spherical symmetry is curious. No WF-like bipolar jet has been identified, but another interferometric observation for confirmation would be desirable. IRAS 22516+0838 is also an AGB star, but it is an interesting object with a similar H$_2$O maser profile to the known WFs.

On the other hand, IRAS 19083+0851 and IRAS 19229+1708 are not considered to be AGB stars because no clear pulsation are observed. These two sources are classified as Group 4 in \citet{2021MNRAS.505.6051J}, not post-AGB (Group 2), and the authors of this paper suggest that these two sources are RSGs. However, one should be careful with the classification of these sources. \citet{2021MNRAS.505.6051J} proposes that IRAS 19083+0851 is an RSG because it shows no pulsation and has a wide OH maser velocity range. However, these properties are also seen in post-AGB stars. It has already been noted above that post-AGB stars do not show pulsation, and a wide range of OH maser velocities such as $>$60~km~s$^{-1}$ has indeed been detected in post-AGB stars \citep{2004ApJS..155..595D}. As for the other source, IRAS 19229+1708, the question also remains whether it shows a pulsation or not, as mentioned in \citep{2004ApJS..155..595D}. In fact, it should be noted that in Gaia DR3, this source is classified as an LPV (long period variable) with a period of 908 days.

With respect to the possibility of the presence of post-AGB stars in the IIIb region, we would like to note that the infrared color of the WF may depend on the direction of observation due to the non-spherically symmetric structure of the WF. In general, the conventional WFs selected only for their H$_2$O maser velocity range above 100~km~s$^{-1}$ are observed from an angle relatively close to the direction of the jet. In contrast, the selection method used in this study may have selected WFs with a larger angle between the line-of-sight direction and the jet direction than WFs selected by the conventional selection method. For example, assuming that the cooler dust component ejected as a jet is in the foreground, obscuring the relatively warmer component in the background, this could possibly explain the presence of post-AGB stars in the IIIb region and the redder infrared color of a conventional WF viewed from a direction closer to the jet. It is beyond the scope of this study and not addressed here, but the relationship between the structure of the dust envelope of the WF, the direction of observation, and the infrared color needs to be confirmed by radiative transfer calculations.

There are 3 Group A sources in regions IV, VIb, and VIII, corresponding to more evolved sources (see the red dots in Figure~\ref{fig:2color}; note that IRAS 19069+0916 is practically located in the IIIb region based on the uncertainty of the 60 $\mu$m flux.). Although these sources may have evolved to the planetary nebula phase in terms of color, it should be noted that the OH maser profiles of these four sources show the double peak profile typically seen in AGB/post-AGB stars (see, Sections~\ref{subsec:18251-1048}, \ref{subsec:18588+0428} and \ref{subsec:19067+0811}). Since there is a large uncertainty in determining the evolutionary stage by mid-infrared colors, the possibility that these sources still have a spherically expanding AGB/post-AGB envelope cannot be completely ruled out. However, note that there are phenomena such that, once a source has evolved into a planetary nebula, it temporarily undergoes AGB-like mass ejections again \citep{2005MNRAS.357.1189C}. In any case, these objects are unlikely to be very young WFs.

\subsection{Possibility of finding by-products} \label{subsec:by-products}

Our initial goal in the present data analysis was to search for WF candidates. However, a closer examination of the spectra for individual sources shows that our selection is likely to select additional interesting sources other than WFs. A good example is K3-35, shown on the right side of Figure~\ref{fig:2color}. This source is known to be a planetary nebula with maser emission \citep[see, e.g.,][]{2001Natur.414..284M}, so its identity is different from that of the WF. However, in common with the WF, the molecular envelope has a deviation from spherical expansion.

We would also like to point out that some sources that meet the condition for velocity screening emit SiO maser lines with a relatively wide linewidth larger than $\sim$15--20 km~s$^{-1}$ (e.g. IRAS 04396+0647, IRAS 15408--5413, IRAS 19083+0851). SiO maser lines in AGB stars usually have a narrow linewidth, typically a few km~s$^{-1}$ \citep[see e.g.,][]{1991A&A...242..211J}. SiO maser lines with a linewidth larger than 10 km~s$^{-1}$ are often detected in red supergiants \citep[RSGs; see e.g.,][]{2010PASJ...62..391D,2012A&A...541A..36V}. RSGs are evolved stars with a large initial mass, and the dynamical structure of their circumstellar molecular envelopes is often more complicated than that of AGB stars \citep[see e.g.,][]{2013ApJS..209...38K}. Although it is arguable whether it is possible to determine RSG status from maser profiles alone, there is no doubt that a non-negligible number of RSGs can be included in the sample for the purpose of WF search.

Then IRAS 19312+1950, although excluded from the present study, also fulfills the velocity condition of our screening \citep[see, e.g.,][]{2011ApJ...728...76N}. The identity of this maser source remained unknown for a long time \citep[see, e.g.,][]{2015PASJ...67...95N,2016ApJ...825...16N}, but recent research has pointed to the possibility of a red nova remnant \citep{2023A&A...669A.121Q}, which is an explosive event that occurs when two stars merge. And importantly, it has been known that a cold molecular envelope is formed after the red nova explosion \citep[see e.g.,][]{2015Natur.520..322K,2021A&A...655A..32K}. The red nova remnants surrounded by a cold molecular envelope are often difficult to distinguish from evolved stars from an observational point of view. However, it is easy to speculate that the molecular envelope of red nova remnants exhibits motion that deviates from a simple spherical expansion, since it is formed by the merger of two stars. Indeed, a radio interferometric observation in molecular lines \citep{2005ApJ...633..282N} and a near-infrared imaging observation \citep{2007A&A...470..957M} revealed the highly complex structure and motion of the envelope of IRAS 19312+1950.

Thus, as described above, many of the objects that may be selected for our screening, even if they are not WFs, are interesting from the point of view of stellar physics and stellar evolution. Therefore, the task of constructing a large list of circumstellar maser sources, regardless of whether they are WFs or not, where the velocity range of the H$_2$O maser line exceeds that of the OH maser line, would be a scientifically significant project.

\section{Summary} \label{sec:summary}
In this study, databases of circumstellar OH (1612 MHz) and H$_2$O (22.235 GHz) maser sources were used to select WF candidates on the basis of a comparison of the velocity ranges of OH and H$_2$O maser emission. By comparing the velocity ranges of the OH and H$_2$O maser emission, it is in principle possible to detect WFs with a smaller velocity range for the H$_2$O maser emission than for the previously known WFs. A cross-check between the 2,195 OH maser sources and the 3,642 H$_2$O maser sources was performed to compare the velocity ranges of OH and H$_2$O maser emission. As part of the selection process, we examined the WISE and 2MASS infrared images and excluded sources with large, extended nebulae, as they are most likely to be YSOs. We also excluded sources that could not be clearly identified with their infrared counterparts due to uncertainties in the coordinate information. In the first stage of selection, the OH and H$_2$O maser velocity ranges were compared for 229 sources, and 41 sources met the velocity criterion.  Then, after a detailed examination of maser line profiles available in the literature, we concluded that the deviation of the H$_2$O maser velocity range was significant for 11 sources among the 41 initially selected sources. The main results are as follows:

\begin{itemize}
\item We examined the IRAS colors of the samples and found that two of the 11 sources with a confirmed H$_2$O maser velocity deviation (IRAS 19069+0916 and IRAS 19319+2214) are in the color region for post-AGB stars. The H$_2$O maser profile of these sources are similar to that of known WFs.

\item Of the 11 selected sources, six sources were located in the color region of the AGB stars. For two of the six sources (IRAS 19422+3506 and IRAS 22516+0838), the H$_2$O maser properties are different from those of typical AGB stars, and confirmation observations with radio interferometry are desired.

\item Among the 11 selected sources, three sources (IRAS 18251-1048, IRAS 18588+0428 and IRAS 19067+0811) were identified in color regions corresponding to more evolved objects. The OH maser profiles of these three sources showed the double peak profile AGB/post-AGB stars. This fact suggests that the infrared colors may depend on the orientation of the object.

\item We also confirmed the possibility that sources exhibiting the velocity deviation of the H$_2$O maser line could include astrophysically interesting sources other than WFs. Such objects could include, for example, peculiar planetary nebulae with maser emission and stellar merger remnants.
\end{itemize}

\begin{acknowledgments}
We acknowledge the science research grants from the China Manned Space Project with No. CMS-CSST-2021-A03, No.CMS-CSST-2021-B01. 
JN acknowledges financial support from the `One hundred top talent program of Sun Yat-sen University' grant no. 71000-18841229. 
YZ are grateful for financial supports from the National Science Foundation of China (NSFC, Grant No. 11973099) and the science research grants from the China Manned Space Project (NO. CMS-CSST-2021-A09 and CMS-CSST-2021-A10).
JJQ thanks for support from the NSFC (No. 12003080), the Guangdong Basic and Applied Basic Research Foundation (No. 2019A1515110588), and the Fundamental Research Funds for the Central Universities (Sun Yat-sen University, No. 22qntd3101).
CHH is supported by the Hong Kong Research Grants Council for GRF research support under grants 17326116 and 17300417. C.-H. Hsia also thanks Q. A. Parker and HKU for the provision of his research post.
\end{acknowledgments}

\appendix

\section{Distribution of circumstellar maser sources} \label{ap:distribution}

\begin{figure}[h]
    \centering
    \includegraphics[width = 0.9\textwidth]{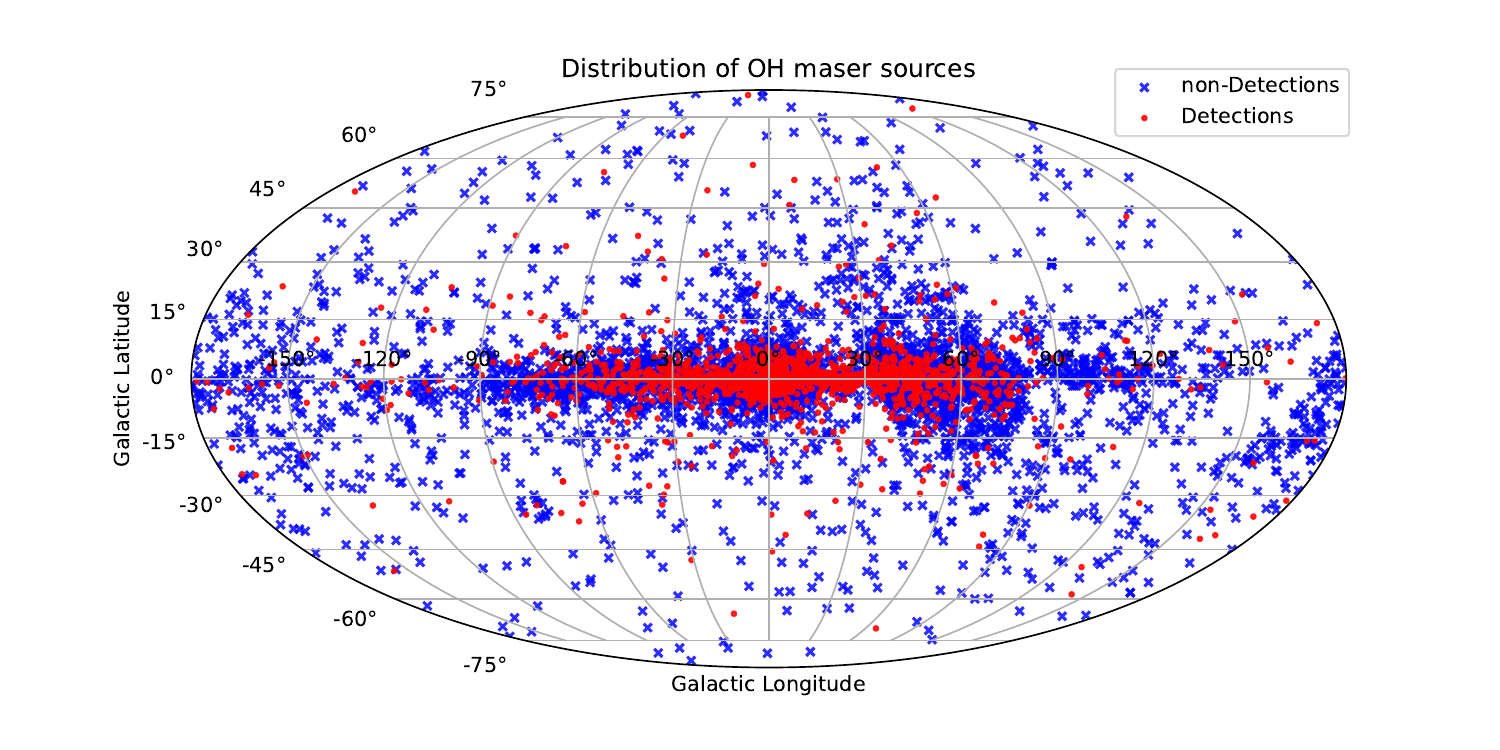}
    \caption{Distribution of 6068 OH maser sources (1612 MHz line) on the galactic coordinate. The data are taken from \citet{2015A&A...582A..68E}. The red filled circles and blue crosses represent detections (2195 sources) and non-detections (3873 sources), respectively.}
    \label{fig:OH-distribution}
\end{figure}

\begin{figure}[h]
    \centering
    \includegraphics[width = 0.9\textwidth]{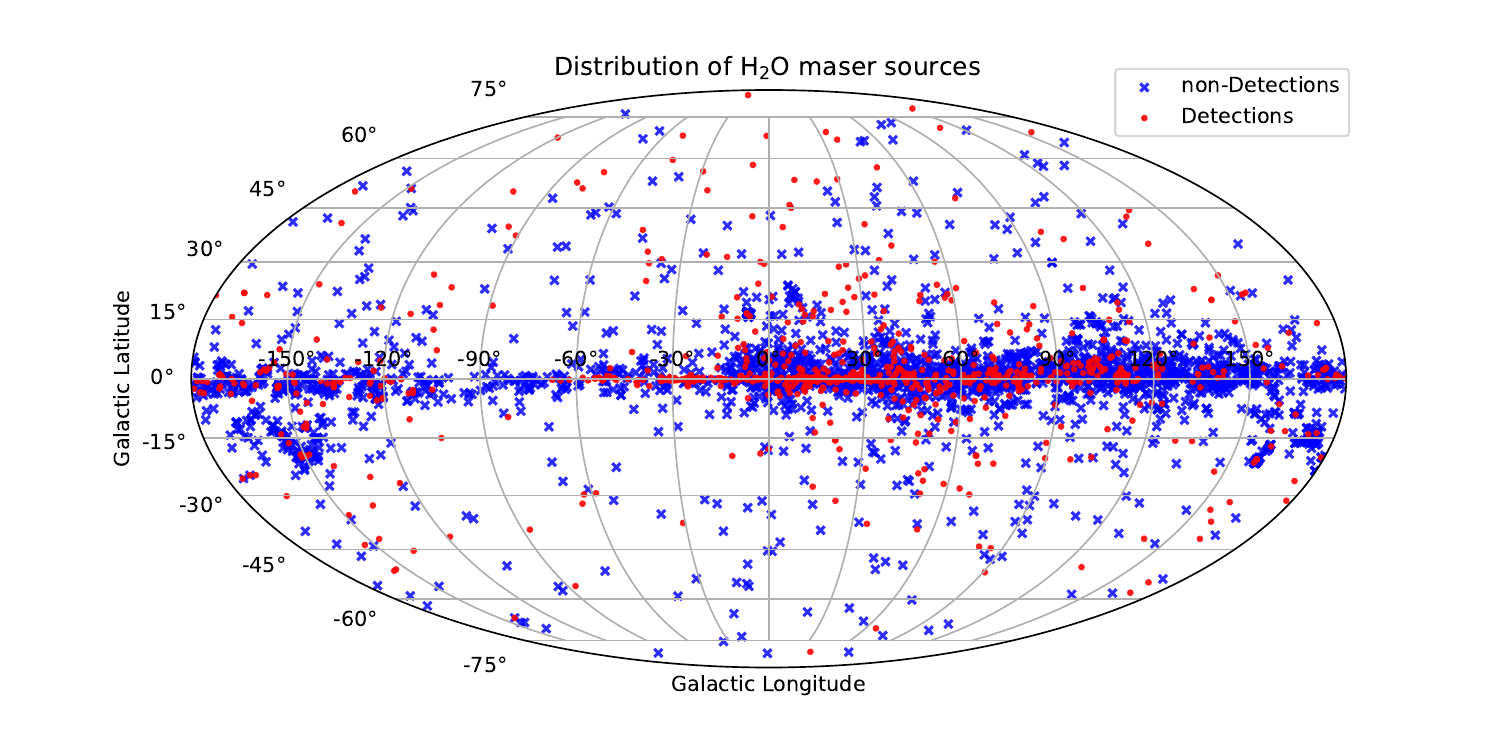}
    \caption{Distribution of 3642 H$_2$O maser sources (22.235 GHz line) on the galactic coordinate. The red filled circles and blue crosses represent detections (890 sources) and non-detections (2752 sources), respectively.}
    \label{fig:H2O-distribution}
\end{figure}

Figures~\ref{fig:OH-distribution}  and \ref{fig:H2O-distribution} show the galactic coordinate distributions of the 1612 MHz OH and 22 GHz H$_2$O maser sources analyzed in this study.

\section{Maser characteristics of Group B sources} \label{ap:indiv}

This section summarizes the characteristics of the maser profiles for each individual source for the objects classified as ``Group B". See Section~\ref{subsec:indiv} for Group B classification criteria.

\subsection{IRAS 04396+0647} \label{subsec:04396+0647}
The OH maser line profile shows a double peak  \citep{1994ApJS...93..549L}, but the velocity difference between the two peaks is somewhat narrower than is typical for AGB stars. Additionally, there is a fast component outside the two peaks, which is not typical of the spherically expanding envelopes of AGB stars. The H$_2$O maser emission is also detected outside the two peaks of the OH maser emission, but the velocity difference between the two peaks is small, about 2.3~km~s$^{-1}$ \citep{2001A&A...368..845V,2017ApJS..232...13C}. Because of the presence of the velocity component of the OH maser emission outside of the two peaks, the velocity range of the H$_2$O maser does not deviate completely from the velocity range of the OH maser emission. The SiO $v=1$ and 2, $J=1-0$ lines have been detected \citep{2017ApJS..232...13C}, with the $v=2$ line showing three peaks and a relatively wider linewidth of about 15~km~s$^{-1}$. The $v=1$ line has a similar linewidth. The radial velocity of the SiO maser lines is almost the same as that of the OH maser line. These maser characteristics are reminiscent of red supergiants, but not of AGB stars or WF candidates.

\subsection{IRAS 05027--2158} \label{subsec:05027-2158}
The OH maser line is detected, but it is very weak, as reported by \citet{2010ARep...54..400R}. It is recorded in the database as having a double peak at velocities of $-$15.9~km~s$^{-1}$ and $-$0.2~km~s$^{-1}$, but there are no supporting spectra found in the literature. The OH 1667 MHz line was detected near $-$30~km~s$^{-1}$ by \citet{2010ARep...54..400R}. The H$_2$O maser line has been observed multiple times \citep{1994PASJ...46..629T,2008PASJ...60.1077S}, exhibiting a highly time-variable profile. According to the information in the database, the velocity difference between the two peaks is 4.6~km~s$^{-1}$, but in the spectra found in the literature, it varies up to about 7~km~s$^{-1}$. The SiO $v=1$ and 2, $J=1-0$ lines have also been detected \citep{1996A&AS..115..117C,2010ApJS..188..209K}. It is a single peak profile with a narrow line width and appears to be a typical SiO maser profile for an AGB star. However, the SiO $v=2$, $J=2-1$ line shows a double peak profile with peaks at approximately $-$25~km~s$^{-1}$ and $-$35~km~s$^{-1}$ \citep{1994A&AS..103..107H}. The OH maser line profile is not clear due to the insufficient signal-to-noise ratio.

\subsection{IRAS 05528+2010} \label{subsec:05528+2010}
The profile of the OH maser line is not the typical double peak, but rather a complex shape with multiple peaks, as reported by \citet{1989A&AS...78..399T}. The H$_2$O maser emission profile is a simple single peak, as reported by \citet{2021ApJ...907...42N}, and it is located just a few~km~s$^{-1}$ outside the velocity range of the OH maser emission. The SiO maser lines have also been observed many times, as reported by, for example, \citet{2010ApJS..188..209K}; the linewidth of the SiO $v=1$ and 2, $J=1-0$ lines is up to 10~km~s$^{-1}$, which is somewhat broad for a typical AGB star. Given that the H$_2$O maser velocity range is small (6.0~km~s$^{-1}$) and the deviation from the OH maser velocity range is also not large, we cannot conclude that the envelope deviates from spherical expansion.

\subsection{IRAS 07445--2613} \label{subsec:07445-2613}
The profile of the OH maser line is the typical double peak of a spherically expanding envelope, as reported by \citet{1985A&AS...59..465S}. The SiO $v=1$ and 2, $J=1-0$ maser lines also show a single-peak profile, which is typical for AGB stars, as reported by \citet{2013AJ....145...22K}. The peak velocity of the SiO maser is roughly equal to the average of the velocities of the OH double peaks. The H$_2$O maser's velocity range deviates slightly from that of the OH maser, with a deviation of only 0.3~km~s$^{-1}$, as reported by \citet{2001A&A...368..845V,2013AJ....145...22K}. The velocity deviation is relatively small, but it is still present. Overall, the evidence is weak to conclude that the envelope deviates from spherical expansion.

\subsection{IRAS 07446--3210} \label{subsec:07446-3210}
Although it is listed in the OH maser source database as having a double-peak profile, we were unable to find any literature containing the spectra. The SiO $v=1$ and 2, $J=1-0$ maser lines and the H$_2$O maser line have been observed \citep{2010ApJS..188..209K}, but there are no special characteristics other than the slightly wide linewidth of the SiO maser (approximately 15~km~s$^{-1}$).

\subsection{IRAS 08357--1013} \label{subsec:08357-1013}
According to the database, the OH maser line has a double-peaked profile with peak velocities of $-$3.9~km~s$^{-1}$ and 39.4~km~s$^{-1}$. However, we couldn't find any literature that confirms these peak velocities through spectra. The spectrum in \citet{1985A&AS...59..465S} confirms the peak on the red-shifted side, but the peak on the blue-shifted side is missing. The SiO maser lines have a relatively wide line width, around 10~km~s$^{-1}$ for the $v=1$, $J=1-0$ line and 7--8~km~s$^{-1}$ for the $v=2$, $J=1-0$ line \citep{2010ApJS..188..209K}.

\subsection{IRAS 13001+0527} \label{subsec:13001+0527}
The database shows that the OH maser line has two peaks at 23.5~km~s$^{-1}$ and 24.0~km~s$^{-1}$. However, the spectrum in  \citet{2003A&A...403L..51E} shows another peak at 11.5~km~s$^{-1}$. Based on the line profile, the peaks at 23.5~km~s$^{-1}$ and 24.0~km~s$^{-1}$ are associated with the receding component of a spherically expanding envelope. Therefore, it is likely that the velocities of approaching and receding components for this source should be around 11.5~km~s$^{-1}$ and 24.0~km~s$^{-1}$. The velocity range of the H$_2$O maser line \citep{2001A&A...368..845V} also falls between these two velocities. The SiO maser lines show the profile of a typical AGB star with a single peak \citep[see, e.g.,][]{2020A&A...642A.213G}. Based on the overall properties of the maser emission lines, it is likely that this source is a typical AGB star, not a WF.

\subsection{IRAS 15060+0947} \label{subsec:15060+0947}
The OH maser profile is that of a typical spherical expanding envelope with double peaks, but there is some fine structure on the blue-shifted side \citep{1994ApJS...93..549L}. The database shows that the minimum and maximum velocities of the H$_2$O maser are $-$14.4~km~s$^{-1}$ and 3.0~km~s$^{-1}$ respectively. However, additional peaks are seen around $-$3~km~s$^{-1}$ \citep{2013ApJ...769...20Y} and 16~km~s$^{-1}$ \citep{2010ApJS..188..209K}. The 16~km~s$^{-1}$ feature is marginal, but if it is a true detection, it can be concluded that the velocity range of the H$_2$O maser deviates from that of the OH maser. Thus, confirmation observation of the H$_2$O maser is desired. The SiO $v=2$, $J=1-0$ line is detected with a narrow peak near $-$12~km~s$^{-1}$ \citep{2010ApJS..188..209K}.

\subsection{IRAS 15255+1944} \label{subsec:15255+1944}
The profile of the OH maser emission line is a typical double peak \citep{2012A&A...537A...5W}. The profile of the H$_2$O maser line shows multiple peaks \citep{2010ApJS..188..209K}, slightly deviating from the velocity range of the OH maser emission (it only exceeds 0.4~km~s$^{-1}$). The profiles of the SiO $v=1$ and 2, $J=1-0$ maser lines \citep{2010ApJS..188..209K} show a single peak which is also within the velocity range of the OH maser emission. Although the velocity range of the H$_2$O maser differs slightly from that of the OH maser, the evidence is weak to determine that the velocity range of the H$_2$O maser line deviates from that of the OH maser line.

\subsection{IRAS 15408--5413} \label{subsec:15408-5413}
According to the information in the database, the profile of the OH maser line is classified as double-peaked with peaks at $-$105.2~km~s$^{-1}$ and $-$101.1~km~s$^{-1}$. However, another peak around $-$60~km~s$^{-1}$ can be seen in the spectrum in \citet{1991A&AS...90..327T}. If the $-$60~km~s$^{-1}$ peak is the red-shifted component of the spherically expanding envelope, the velocity range of the H$_2$O maser emission is within that of the OH maser emission. Considering the $-$60~km~s$^{-1}$ peak, the expansion velocity of the spherically expanding envelope should be about 20~km~s$^{-1}$.  The H$_2$O maser spectra in the literature \citep{2011MNRAS.416.1764W,2015A&A...578A.119G} agree with the velocity information in the database. The SiO $v=1$, $J=2-1$ maser line was detected \citep{1990A&A...233..477L}. The linewidth of the SiO maser line is somewhat wide (about 25~km~s$^{-1}$), which is more RSG-like than AGB-like. 


\subsection{IRAS 17343+1052} \label{subsec:17343+1052}
The OH maser line profile is a typical double peak \citep{1994ApJS...93..549L}. The SiO $v=1$ and 2, $J=1-0$ maser lines are detected and the line profile is a typical single peak \citep{2017ApJS..232...13C}. The profile of the H$_2$O maser line \citep{2013ApJ...769...20Y, 2017ApJS..232...13C} has a shape that extends over the velocity range of the OH maser emission. According to the information in the database, the velocity range of the H$_2$O maser emission exceeds that of the OH maser emission by only about 0.4~km~s$^{-1}$. We have not confirmed the possibility of high velocity components in the H$_2$O maser spectra. The evidence is weak to conclude that the velocity range of the H$_2$O maser deviates from that of the OH maser.

\subsection{IRAS 17392--3020} \label{subsec:17392-3020}
According to the information in the database, the profile of the OH maser line is double-peaked with peaks at $-$24.9~km~s$^{-1}$ and 18.5~km~s$^{-1}$. The OH maser spectra in the literature \citep{2004ApJS..155..595D} show another peak at $-$10~km~s$^{-1}$. However, the $-$10~km~s$^{-1}$ peak has an unusually narrow line width (probably 1 channel), so it is possible that it is a spurious line. The profile of the H$_2$O maser line \citep{2007ApJ...658.1096D} is a double peak with peaks at similar velocities to the OH maser line. No reports of SiO maser detected. We believe the evidence is weak to conclude that the velocity range of the H$_2$O maser deviates from that of the OH maser.

\subsection{IRAS 17540-1919} \label{subsec:17540-1919}
The database shows that the OH maser line has two peaks at 56.4~km~s$^{-1}$ and 64.0~km~s$^{-1}$. Although these peaks can be seen in the spectrum in \citet{ 1989A&AS...78..399T}, the red shifted peak is faint. Judging from the spectrum given in \citet{ 2017ApJS..232...13C}, the profile of the H$_2$O maser line is single-peaked, and the peak velocity is consistent with the information in the database. The SiO $v=1$, $J=1-0$ line is detected \citep{ 2017ApJS..232...13C}; the profile of the SiO maser line is a typical single peak of an AGB star, and the peak velocity is about the same as that of the H$_2$O maser. The H$_2$O maser emission deviates from the velocity range of the OH maser line, but only by a small amount (1.4~km~s$^{-1}$).

\subsection{IRAS 17580--3111} \label{subsec:17580-3111}
In the OH maser spectra presented in the literature \citep{ 1991A&AS...90..327T,1994A&AS..103..301H} four peaks can be identified in the velocity range from about 0~km~s$^{-1}$ to 30~km~s$^{-1}$. According to the information in the database, the velocity range of the H$_2$O maser line is from 0.4~km~s$^{-1}$ to 21.3~km~s$^{-1}$. However, in the spectra presented in the literature \citep{2007A&A...467.1085S,2008AJ....135.2074G}, a strong peak was observed only around 21~km~s$^{-1}$, but no emission was observed around 0~km~s$^{-1}$. The spectrum in \citet{2007A&A...467.1085S} shows the possibility of a weak emission around 47~km~s$^{-1}$, but it is marginal. There are no reports of SiO maser lines being detected.

\subsection{IRAS 18009--2019} \label{subsec:18009-2019}
The profile of the OH maser line presented in the literature \citep{1989A&AS...78..399T,2001A&A...368..845V} is a typical double peak. The velocity range also matches the information in the database. The velocity range of the H$_2$O maser line given in the database is $-$1.5~km~s$^{-1}$ to 21.0~km~s$^{-1}$, but the spectra found in the literature \citep{1994PASJ...46..629T,2010ApJS..188..209K} did not confirm the blue-shifted end (i.e. $-$1.5~km~s$^{-1}$). The OH maser velocity range given in the database is from 3.0~km~s$^{-1}$ to 26.0~km~s$^{-1}$, but the H$_2$O maser spectra in the literature do not clearly show any velocity excess. The SiO $v=1$ and 2, $J=1-0$ lines are detected, and the line profile is a typical single peak with a peak velocity near 10~km~s$^{-1}$ \citep{1977MNRAS.180..415B,1996A&AS..115..117C,2010ApJS..188..209K}.

\subsection{IRAS 18034--2441} \label{subsec:18034-2441}
Spectra confirming the OH maser line profile cannot be found in the literature. According to the spectrum of the H$_2$O maser line in \citet{2017ApJS..232...13C}, the line profile is a single peak with a peak velocity near $-$13~km~s$^{-1}$, similar to the velocity in the database. No velocity components other than the $-$13~km~s$^{-1}$ peak can be identified. The SiO $v=1$ and 2, $J=1-0$ lines have been observed twice with an interval of more than 15 years, but the two profiles differ \citep{2004PASJ...56..765D,2017ApJS..232...13C}. The spectrum reported by \citep{2017ApJS..232...13C} shows a double peak profile with peaks at about 2~km~s$^{-1}$ and 8~km~s$^{-1}$, while that reported by \citep{2004PASJ...56..765D} shows a single peak profile with a peak at about 8~km~s$^{-1}$. Considering that the peak velocities of the SiO maser lines are close to the systemic velocity of the source, the feature of the H$_2$O maser line presumably traces the approaching side of a spherically expanding envelope. The deviation of the H$_2$O maser line from the velocity range of the OH maser line is small (0.5~km~s$^{-1}$).

\subsection{IRAS 18050--0518} \label{subsec:18050-0518}
We could not find any literature with spectra of the OH maser line. The H$_2$O maser spectrum has a peak near $-$28~km~s$^{-1}$ and shows a relatively broad emission from about $-$32~km~s$^{-1}$ to $-$21~km~s$^{-1}$ \citep{2013ApJ...769...20Y}. The SiO $v=1$ and 2, $J=1-0$ lines are detected, both peaking at $-$32~km~s$^{-1}$ \citep{2012PASJ...64....4D}. The profiles of the SiO maser lines are a typical single peak of AGB stars. According to the information in the database, the deviation of the H$_2$O maser emission line from the velocity range of the OH maser line is small (0.3~km~s$^{-1}$), and the profile of the H$_2$O maser line does not indicate the possible presence of a high-velocity component.

\subsection{IRAS 18052--2016} \label{subsec:18052-2016}
The OH maser line profile, as reported in various studies such as \citep{1989A&AS...78..399T,1994A&A...287..479B,2004ApJS..155..595D,2012A&A...537A...5W}, displays the typical double peak of a spherically expanding envelope. According to the information in the database, the peak velocity of the blue-shifted component of the OH maser line is 23.0~km~s$^{-1}$. Additionally, the H$_2$O maser profile exhibits a weak extended wing component near this OH peak velocity, as reported in \citep{2007ApJ...658.1096D,2014ApJS..211...15Y}. However, it is unclear from the spectra whether the H$_2$O maser emission exists outside the velocity range of the OH maser line. We note that the OH maser spectrum in \citet{2004ApJS..155..595D} reveals that the OH maser emission of another evolved star (b298) is also detected in the same spectrum, and the velocity of the red-shifted peak of b298 is very close to the blue-shifted peak of IRAS 18052--2016 (with a velocity separation of about 5~km~s$^{-1}$). Therefore, it is possible that the velocity excess of H$_2$O may be due to the emission of b298. The SiO $v=1$ and 2, $J=1-0$ and $^{29}$SiO $v=1$, $J=1-0$ lines have been detected according to the reference \citep{2014ApJS..211...15Y}. These lines exhibit a typical single-peak profile of an AGB star. The peak velocity is around 53~km~s$^{-1}$, which is roughly the average of the two peaks of the OH maser line. Due to the lack of sensitivity in the observations, the velocity of the wing component of the H$_2$O maser is not clearly known, making it difficult to completely rule out the possibility of the velocity deviation of H$_2$O maser emission.

\subsection{IRAS 18085+0752} \label{subsec:18085+0752}
The profile of the OH maser line is a typical double peak \citep{1994ApJS...93..549L}. The H$_2$O maser line has been observed twice. The spectrum in \citep{2017ApJS..232...13C} shows a simple single peak profile with a peak near $-$68~km~s$^{-1}$, while the spectrum in \citep{2013ApJ...769...20Y} shows a peak near $-$64~km~s$^{-1}$ in addition to the $-$68~km~s$^{-1}$ peak. No other possible velocity components can be seen in the H$_2$O spectrum. The SiO $v=1$ and 2, $J=1-0$ lines are observed and only the $v=2$ line is detected \citep{2010PASJ...62..525D}. The fact that only the $v=2$ line is detected suggests that this is an AGB star relatively close to the post-AGB phase \citep{2007ApJ...669..446N}. The deviation of the H$_2$O maser line from the OH maser velocity range is small (0.6~km~s$^{-1}$).

\subsection{IRAS 18348--0526} \label{subsec:18348-0526}
The OH maser line profile is a typical double peak \citep{1989A&AS...78..399T, 2012A&A...537A...5W}. The OH 1665 MHz and 1667 MHz lines are also detected and show double peaked line profiles with peak velocities nearly the same as the OH 1612 MHz line \citep{1989A&AS...78..399T}. In the H$_2$O maser spectra, two peaks can be seen at about 16~km~s$^{-1}$ and 36~km~s$^{-1}$ \citep{1994PASJ...46..629T,2007ApJ...669..446N}. The information in the database indicates that the maximum velocity of the H$_2$O maser exceeds the redshifted peak of the OH maser line by 3.0~km~s$^{-1}$, but no spectra were available in the literature to confirm this deviation. The SiO $v=0$, 1, 2 and 3, $J=1-0$ lines and the $^{29}$SiO $v=0$, $J=1-0$ lines are detected \citep{2007ApJ...669..446N}. The $v=0$ lines are thought to be thermal lines, but the line widths are not much different from the other SiO maser lines.

\subsection{IRAS 18436+4334} \label{subsec:18436+4334}
The OH 1612 MHz maser line is a typical double peak \citep{1989A&AS...78..399T,2001A&A...368..845V}. The OH 1665 MHz and 1667 MHz lines are also detected, but the line profiles are different from the OH 1612 MHz line, with velocity components of these lines closer to the system velocity. The H$_2$O maser spectrum shows a strong peak at about $-$22~km~s$^{-1}$ \citep{2001A&A...368..845V,2010ApJS..188..209K}, but no fast components that would deviate from the velocity range of the OH maser line can be identified. The SiO $v=1$ and 2, $J=1-0$ lines and the SiO $v=1$, $J=2-1$ and $3-2$ lines are detected \citep{1998ApJS..115..277C,2009ApJS..181..421C,2010ApJS..188..209K}. The line profiles are typical single peaks for all lines. The deviation of the H$_2$O maser line from the velocity range of the OH maser line is small (0.8~km~s$^{-1}$).

\subsection{IRAS 18494--0130} \label{subsec:18494-0130}
The profile of the OH maser line is a typical double peak \citep{ 1989A&AS...78..399T}. The H$_2$O maser line is strongly detected at velocities close to the blue-shifted peak of the OH maser line \citep{ 2013ApJ...769...20Y}. The emission can also be weakly detected in the red-shifted peak of the OH maser line, although this is marginal. The SiO $v=1$ and 2, $J=1-0$ line is detected near 78~km~s$^{-1}$, and the line profile is a typical single peak \citep{1991A&A...242..211J,2004PASJ...56..765D}. According to the information in the database, the deviation of the H$_2$O maser line from the velocity range of the OH maser line is only 0.8~km~s$^{-1}$.

\subsection{IRAS 19395+1949} \label{subsec:19395+1949}
The OH maser profile is double peaked in the database classification. However, we can speculate that both peaks are emissions associated with the approaching side of a spherically expanding envelope, according to the maser spectra in the literature \citep{1990ApJ...362..634L,2013ApJ...769...20Y,2017ApJS..232...13C}. The reason is that the velocity interval between the two peaks ($\sim$3.2~km~s$^{-1}$) is too narrow for an AGB envelope \citep{1990ApJ...362..634L}, and both peaks are on the blue-shifted side of the SiO maser line, which has a peak at about 44~km~s$^{-1}$ \citep{2017ApJS..232...13C}. According to the information in the database, the velocity range of the H$_2$O maser line is 28.0~km~s$^{-1}$ to 56.0~km~s$^{-1}$, which agrees with the spectrum in the literature \citep{2013ApJ...769...20Y}. The red-shifted end of the velocity range of the H$_2$O maser is 56~km~s$^{-1}$, which is only 31~km~s$^{-1}$ away from the blue-shifted end of the velocity range (25~km~s$^{-1}$) of the OH maser line. This difference in velocity can be converted to an expansion velocity of 15.5~km~s$^{-1}$, which is within the range of expansion velocities of normal AGB envelopes.

\subsection{IRAS 19466+2751} \label{subsec:19466+2751}
The OH maser spectrum from the literature shows four marginal peaks \citep{1994ApJS...93..549L}, and their velocities are not inconsistent with the information in the database. However, it is difficult to tell from the spectra of \citet{1994ApJS...93..549L} whether the OH maser detection is true because of the high noise level. The detection of the H$_2$O maser is clearly confirmed by the spectra in the literature \citep{2001A&A...368..845V}, and the velocity is consistent with the information in the database. If we believe the velocity range of the OH maser line given in the database, the velocity range of the H$_2$O maser emission would deviate from that of the OH maser emission. However, the signal-to-noise ratio of the OH maser spectrum is not good.

\subsection{IRAS 20015+3019} \label{subsec:20015+3019}
The OH maser profile is classified as double peaked in the database. In the spectra given in the literature, peaks are found at corresponding velocities \citep{1993ApJS...89..189C}. However, the narrow velocity spacing between the two peaks (1.8~km~s$^{-1}$) makes it unlikely that they are emitted from the receding and approaching sides of a spherically expanding envelope. Presumably, both were emitted from either the receding or the approaching side. According to the spectra in the literature \citep{1996A&AS..116..117E}, the H$_2$O maser line was detected at velocities consistent with the information in the database. The difference between the velocity at the blue-shifted feature of the OH maser line (5.5~km~s$^{-1}$) and the velocity at the red-shifted end of the H$_2$O maser (20.1~km~s$^{-1}$) is 14.6~km~s$^{-1}$, which is not above the expansion velocity of the normal AGB envelope.

\subsection{IRAS 20440+0412} \label{subsec:20440+0412}
The database information classifies the OH maser profile of this source as double peaked, and the peak velocities are consistent with the spectra found in the literature \citep{1989A&AS...78..399T,1990ApJ...362..634L}. However, the velocities of the two peaks are close together at $-$54.8~km~s$^{-1}$ and $-$49.9~km~s$^{-1}$, making it unlikely that they are peaks associated with the receding and approaching sides of a spherically expanding envelope. It is speculated that both peaks are associated with either the receding or the approaching side. In the literature spectrum \citep{2013ApJ...769...20Y,2017ApJS..232...13C}, the H$_2$O maser profile is a single peak with a peak velocity near $-$50~km~s$^{-1}$. The SiO $v=1$ and 2, $J=1-0$ lines are detected \citep{2017ApJS..232...13C}. The SiO maser profile is a single peak, similar to the H$_2$O maser line, with a peak around 50~km~s$^{-1}$. This is a relatively rare case where maser lines from three molecules, OH, H$_2$O, and SiO, are detected at approximately the same velocity. However, the H$_2$O maser profile shows no signs of a jet, such as high-velocity components or multiple peaks.

\subsection{IRAS 20547+0247} \label{subsec:20547+0247}
The OH maser profile in the literature spectrum is double-peaked \citep{1990A&A...233..112S,1993ApJS...89..189C}. The velocities of the peaks seen in the spectrum also match the information in the database. However, the velocity interval between the peaks is narrow, less than 5~km~s$^{-1}$, and both are probably associated with either the receding or approaching side of the spherically expanding envelope. In the OH 1667 MHz spectrum \citep{1990A&A...233..112S}, a weak emission feature appears to be detected around $-$63~km~s$^{-1}$. If the detection of the $-$63~km~s$^{-1}$ component of the OH 1667 MHz line is true, then the two peaks of the OH 1612 MHz line would be associated with the approaching side of the envelope. The H$_2$O maser profile is a simple single peak with a peak velocity around $-$75~km~s$^{-1}$, as given in the database \citep{1987A&A...173..263Z}.

\subsection{IRAS 22556+5833} \label{subsec:22556+5833}
According to the information in the database, the OH maser profile is classified as double-peaked with peak velocities at $-$67.0~km~s$^{-1}$ and $-$59.0~km~s$^{-1}$. However, another peak is also seen near $-$45~km~s$^{-1}$ in the spectrum of \citet{2017ARep...61...16A}. In addition, strong temporal variations in intensity are seen on a timescale of a few years. The profile of the H$_2$O maser line has a complex shape with multiple peaks over a wide velocity range \citep[about $-$62~km~s$^{-1}$ to $-$40~km~s$^{-1}$ or so,][]{1994PASJ...46..629T,1993A&AS...98..589W,2010ApJS..188..209K}. The SiO $v=1$ and 2, $J=1-0$ lines were detected \citep{2010ApJS..188..209K}. The SiO maser profile is a single peak. Given the presence of the $-$45~km~s$^{-1}$ component, the velocity range of the H$_2$O maser does not deviate from that of the OH maser.

\section{Maser characteristics of other type sources} \label{ap:other}

In this section, we summarize the maser properties of two of the 41 sources selected in the process described in Section~\ref{sec:Methodology} that have been identified as non-AGB and post-AGB star types.

\subsection{IRAS 17443--2949} \label{subsec:17443-2949}
This source is classified as a planetary nebula in the SIMBAD object type \citep[see, e.g.,][]{1994AN....315...63K}. 
The information in the database shows that the profile of the OH maser line is double-peaked with peak velocities at $-$16.7~km~s$^{-1}$ and $-$4.8~km~s$^{-1}$. Although the double-peak profile can be recognized in the spectra \citep{ 2008AJ....135.2074G}, the red-shifted side peak is weak. Most velocity components of the H$_2$O maser emission \citep{2007A&A...467.1085S,2008AJ....135.2074G} fall within the velocity range of the OH maser emission. However, the H$_2$O maser line shows an isolated velocity component in the vicinity of 1~km~s$^{-1}$. This velocity component deviates from the velocity range of the OH maser line by 5.71~km~s$^{-1}$. There are no reports of SiO maser detection.


\subsection{IRAS 19255+2123} \label{subsec:19255+2123}
 This object is classified as a young planetary nebula \citep[K 3-35, see, e.g.,][]{1990AJ....100..485V,2001Natur.414..284M}.
 According to information in the database, the OH maser profile is classified as double-peaked, with peak velocities of $-$3.0~km~s$^{-1}$ and 9.0~km~s$^{-1}$. In the spectrum given in \citet{1985A&A...148..344E}, the OH maser line appears to be weakly detected in the vicinity of 20~km~s$^{-1}$ as well. The H$_2$O maser spectrum shows about three peaks at velocities around 20~km~s$^{-1}$ to 25~km~s$^{-1}$ \citep{1985A&A...148..344E}. The velocities at which the H$_2$O maser line was detected are clearly outside the velocity range of the OH maser line. No SiO maser emission lines were detected. A VLBI observation in the H$_2$O maser line has been made by \citet{2001Natur.414..284M}. The H$_2$O maser line has been detected from a bipolar structure according to the VLBI map of \citet{2001Natur.414..284M}.  The location where the H$_2$O maser line was detected is as far as 5000 AU from the central star.

\section{List of YSO candidates with fast outflows} \label{ap:YSO}

 Some of the sources classified as YSO candidates mentioned in Section~\ref{subsec:h2o-catalog} have H$_2$O maser velocity ranges exceeding 100~km~s$^{-1}$. For the purpose of avoiding confusion with WF, these sources are summarized in Table~\ref{tab:YSOs}.

\begin{deluxetable*}{lrrrrr}
\tabletypesize{\scriptsize}
\tablecolumns{6}
\tablecaption{YSO candidates with fast outflows \label{tab:YSOs}}
\tablehead{
\colhead{Source name} & \multicolumn{2}{c}{Position} & \colhead{$V_{\rm min}$ (H$_2$O)} &  \colhead{$V_{\rm max}$ (H$_2$O)} & \colhead{$\Delta V_{\rm H_2O}$} \\
\colhead{} & \colhead{R.A.} & \colhead{Dec.} & \colhead{} & \colhead{} & \colhead{}  \\
\colhead{} & \multicolumn{2}{c}{(J2000.0)} & \colhead{(km s$^{-1}$)} & \colhead{(km s$^{-1}$)} & \colhead{(km s$^{-1}$)}
}
\startdata
G21.80$-$0.12 & 18 31 21.3 & $-$09 56 56  & $-$53.1     & 164.7      & 217.8\\
NGC 7129 FIR & 21 43 01.3  & +66 03 37  & $-$71.5     & 46.0         & 117.5\\
IRAS 21519+5613 & 21 53 39.2 & +56 27 45 & $-$100.5    & $-$0.5       & 100.0\\
IRAS 20081+3122 & 20 10 09.2  & +31 31 34 & $-$59.0       & 63.5       & 122.5\\
IRAS 15452$-$5459 & 15 49 11.3 & $-$55 08 51  & $-$83.0       & 22.0         & 105.0\\
IRAS 19078+0901 & 19 10 15.6 & +09 06 08    & $-$272.5    & 268.5      & 541.0\\
IRAS 19213+1424 & 19 23 40.0   & +14 31 06  & $-$30.0       & 170.0        & 200.0\\
Orion$-$A & 05 35 14.5  & $-$05 22 21  & $-$119.0      & 67.5       & 186.5\\
IRAS 02232+6138 & 02 27 04.1   & +61 52 22 & $-$117.0      & 6.0          & 123.0\\
NGC 6334I & 17 20 53.0   & $-$35 46 57 & $-$71.0       & 50.0         & 121.0\\
IRAS 02219+6152 & 02 25 40.5  & +62 05 52  & $-$100.5    & 18.0         & 118.5\\
IRAS 18162$-$2048 & 18 19 11.9 & $-$20 47 34 & $-$93.5     & 19.0         & 112.5\\
Orion$-$S & 05 35 12.5  & $-$05 24 11  & $-$57.0       & 54.5       & 111.5\\
NGC 7538 IRS 11 & 23 13 44.8 & +61 26 53 & $-$153.5    & $-$46.5      & 107.0\\
IRAS 22543+6145 & 22 56 17.9 & +62 01 46  & $-$67.0       & 39.5       & 106.5\\
\enddata
\end{deluxetable*}

\section{Reference of 22.235 GHz H$_2$O and 1612 MHz OH maser observations for known WFs} \label{ap:known WF masers}

Table~\ref{tab:maser spectra of known WFs} summarizes the papers with 22.235 GHz H$_2$O maser and 1612 MHz OH maser spectra of known WFs. See Appendix~\ref{ap:code} for the reference codes given in Table~\ref{tab:maser spectra of known WFs}.

\begin{deluxetable*}{lll}
\tabletypesize{\scriptsize}
\tablecolumns{3}
\tablecaption{Maser spectra references for known WFs. \label{tab:maser spectra of known WFs}}
\tablehead{
\colhead{Source name} & \colhead{22.235 GHz H$_2$O lines} & \colhead{1612 MHz OH lines}
}
\startdata
IRAS 15103$-$5754 & SUA09 & LIN96\\
IRAS 15445$-$5449 & DEA07, TAF14 & DEA04, SEV02\\
IRAS 15544$-$5332 & DEA07, GOM17 & DEA04, SEV02\\
IRAS 16342$-$3814 & CLA09, LIK92 & LIN96, WOL12\\
IRAS 16552$-$3050 & SUA07, SUA08 & \\
IRAS 17291$-$2147 & GOM15A, GOM17 & \\
IRAS 18043$-$2116 & DEA07, TAF14, USC23 & DEA04\\
IRAS 18113$-$2503 & GOM15A, YOO14 & \\
IRAS 18139$-$1816 & BOB07, ENG02 & AMI11, BOB05\\
IRAS 18286$-$0959 & TAF14, YUN14 & CHE17\\
IRAS 18450$-$0148 & DEA07, LIK92 & DEA04, WOL12\\
IRAS 18455+0448 & YUN13, YUN14 & LEW01, CHE93\\
IRAS 18460$-$0151 & IMA13 & IMA13, WOL12\\
IRAS 18596+0315 & DEA07, ENG02, GOM15A, GOM17 & AMI11, DEA04, GOM17\\
IRAS 19134+2131 & GOM15A, LIK92, YUN14 & \\
IRAS 19190+1102 & DAY10, GOM15A & LIK89\\
\enddata
\end{deluxetable*}

\section{Machine-readable catalog of H$_2$O maser sources} \label{ap:h2o-catalog}

A catalog of the 890 H$_2$O maser sources used in Section~\ref{subsec:h2o-catalog} is attached to this paper as an electronic table. The electronic table consists of 11 columns (see,  Table~\ref{tab:h2o catalog}). The 11 columns, in order from the beginning, give the serial number, the source name, the right ascension used in the radio observations, the declination used in the radio observations, the right ascension of the WISE counterpart, the declination of the WISE counterpart, the minimum velocity (low veloc), the maximum velocity (high veloc), the velocity range (veloc width), the image inspection results (image-inspec), and literature information are given. The equinox of right ascension and declination is J2000.0. The unit for radial velocity is ``km~s$^{-1}$''. If multiple observations have been made for the same source, the maximum and minimum velocities are calculated from all previous observations (see Section~\ref{subsec:h2o-catalog}). The values of right ascension and declination used for the radio observations were obtained from the references marked with "*". Information on the references corresponding to the literature codes is given in Appendix~\ref{ap:code}. The results of the infrared image inspection fall into three main categories. In the``image-insepc" column, ``point source" indicates that a point source with an evolved star-like infrared color is found in the vicinity of the radio observation location, ``YSO" indicates that an extended infrared nebulosity is seen, and ``crowded region" indicates that many stars are crowded together and no single counterpart can be identified. Additionally, ``!" indicates that no corresponding infrared objects were found in the vicinity of the location where the radio observation was made.

\begin{deluxetable*}{lll}
\tabletypesize{\scriptsize}
\tablecolumns{3}
\tablecaption{Descriptive table of H$_2$O maser source catalog \label{tab:h2o catalog}}
\tablehead{
\colhead{Column} & \colhead{Unit/Format} &  \colhead{description}
}
\startdata
index &  & Row number of the table.\\
source name &  & Name of the observed source. \\
R.A.(J2000) & hh mm ss.s & Right ascension of the observed source. (The rows are sorted by the right ascensions.) \\
DEC(J2000) & dd mm ss.s & Declination of the observed source. \\
WISE R.A. & hh mm ss.s & Right ascension of the WISE counterpart. \\
WISE DEC & dd mm ss.s & Declination of the WISE counterpart. \\
low veloc & km s$^{-1}$ & Minimum H$_2$O maser velocity of the source among all the observations.\\
high veloc & km s$^{-1}$ & Maximum H$_2$O maser velocity of the source among all the observations.\\
veloc width & km s$^{-1}$ & Velocity difference between the maximum velocity and minimum velocity of the source.\\
image-inspec &  & Image inspection result. "!" means no corresponding WISE object was found in the vicinity of the source.\\
reference &  & References of the observations. The observed positions are cited from\\
 &  & the reference with "*" sign.\\
\enddata
\end{deluxetable*}

\section{Reference codes used in Tables~1, 5 and machine-readable table} \label{ap:code}

The correspondence between the literature and the reference codes given in the last column of Table \ref{tab:velocity} is as follows.
AMI10:\citet{2010A&A...509A..26A},
AMI11:\cite{2011A&A...532A.149A},
BOB05:\cite{2005ApJ...627L..45B},
BOB07:\cite{2007ApJ...665..680B},
CHE17:\cite{2017MNRAS.468.3602C},
CHE93:\cite{1993ApJS...89..189C},
CHO10:\citet{2010ApJ...719..126C},
CHO12:\citet{2012AJ....144..129C},
CHO14:\citet{2014JKAS...47..293C},
CHO17:\citet{2017ApJS..232...13C},
CLA09:\cite{2009ApJ...691..219C},
COH06:\citet{2006MNRAS.369..189C},
DAY10:\cite{2010ApJ...713..986D}.
DEA07:\citet{2007ApJ...658.1096D},
DEG04C:\citet{2004PASJ...56.1083D},
DEG07A:\citet{2007ApJ...664.1130D},
DEG10A:\citet{2010PASJ...62..391D},
DEG10B:\citet{2010PASJ...62..525D},
DEG89:\citet{1989MNRAS.239..825D},
ENG02:\cite{2002A&A...388..252E},
ETO17:\citet{2017MNRAS.468.1703E},
FOK12:\citet{2012ApJ...760...65F},
GOM08:\citet{2008AJ....135.2074G},
GOM15A:\citet{2015A&A...578A.119G},
GOM17:\citet{2017MNRAS.468.2081G},
GOM90:\citet{1990RMxAA..20...55G},
GOM94:\citet{1994RMxAA..28...97G},
GRE04:\citet{2004ApJ...601..921D},
GVA14:\citet{2014MNRAS.437..843G},
IMA13:\cite{2013ApJ...773..182I},
KIM10:\citet{2010ApJS..188..209K},
KIM13:\citet{2013AJ....145...22K},
LEP76:\citet{1976A&A....48..269L},
LEP77:\citet{1977A&A....56..219L},
LEW01:\cite{2001ApJ...548L..77L},
LIK89:\cite{1989ApJ...344..350L}.
LIN96:\cite{1996A&AS..119..459T},
NAK00:\citet{2000PASJ...52L..43N},
NAK03B:\citet{2003PASJ...55..229N},
NAK06:\citet{2006ApJ...647L.139N},
NAK07:\citet{2007ApJ...669..446N},
NAK16:\citet{2016PASJ...68...78N},
NAKA93:\citet{1993PASJ...45..179N},
NEU17:\citet{2017ApJ...843...94N},
NEU21:\citet{2021ApJ...907...42N},
OHN13:\citet{2013A&A...559A.120O},
PAS06:\citet{2006A&AT...25..399P},
SEV02:\cite{2002AJ....123.2772S},
SJO02:\citet{2002A&A...391..967S},
SUA07:\citet{2007A&A...467.1085S},
SUA08: \cite{2008ApJ...689..430S},
SUA09:\citet{2009A&A...505..217S},
SUN07:\citet{2007PASJ...59.1185S},
TAF14:\cite{2014A&A...562L...9T},
TAK01:\citet{2001PASJ...53..517T},
TAK94:\citet{1994PASJ...46..629T},
USC23:\citet{2023ApJ...948...17U}
VAL01:\citet{2001A&A...368..845V},
VLE14:\citet{2014A&A...569A..92V},
WAL11:\citet{2011MNRAS.416.1764W},
WAL14:\citet{2014MNRAS.442.2240W},
YOO14:\citet{2014ApJS..211...15Y},
YUN13:\citet{2013ApJ...769...20Y},
YUN14:\citet{2014ApJ...794...81Y},
ZUC87:\citet{1987A&A...173..263Z}.


\bibliography{export-bibtex}{}
\bibliographystyle{aasjournal}



\end{CJK*}
\end{document}